\documentclass[12pt,preprint]{aastex}

\newcommand{\be}{\begin{equation}}
\newcommand{\ee}{\end{equation}} 
\newcommand{\simless}{\lower.5ex\hbox{$\; \buildrel < \over \sim\;$}}
\newcommand{\simgreat}{\lower.5ex\hbox{$\; \buildrel > \over \sim\;$}} 
\newcommand{\fone}{{ {\cal F}_1 }} 
\newcommand{\rclust}{{ R_{\rm c} }} 
\newcommand{\uvexp}{{\langle L_{UV} \rangle_\ast }} 
\newcommand{\sigbar}{{ \langle \sigma \rangle_{UV} }} 
\newcommand{\ncolumn}{{ N_{\rm col} }}  
\newcommand{\mubar}{{\langle \mu \rangle }}  

\begin{document}

\title{UV Radiation Fields Produced by Young Embedded Star Clusters} 

\author{Marco Fatuzzo$^1$ and Fred C. Adams$^{2,3}$ } 
 
\affil{$^1$Physics Department, Xavier University, Cincinnati, OH 45207} 

\affil{$^2$Michigan Center for Theoretical Physics, University of Michigan \\
Physics Department, Ann Arbor, MI 48109}  

\affil{$^3$Astronomy Department, University of Michigan, Ann Arbor, MI 48109}

\email{fatuzzo@xavier.edu, fca@umich.edu} 

\begin{abstract} 

A large fraction of stars form within young embedded clusters, and
these environments produce a substantial ultraviolet (UV) background
radiation field, which can provide feedback on the star formation
process. To assess the possible effects of young stellar clusters on
the formation of their constituent stars and planets, this paper
constructs the expected radiation fields produced by these clusters.
We include both the observed distribution of cluster sizes $N$ in the
solar neighborhood and an extended distribution that includes clusters
with larger $N$. The paper presents distributions of the FUV and EUV
luminosities for clusters with given stellar membership $N$,
distributions of FUV and EUV luminosity convolved over the expected
distribution of cluster sizes $N$, and the corresponding distributions
of FUV and EUV fluxes. These flux distributions are calculated both
with and without the effects of extinction.  Finally, we consider the
effects of variations in the stellar initial mass function on these
radiation fields. Taken together, these results specify the
distributions of radiation environments that forming solar systems are
expected to experience.

\end{abstract}


\section{INTRODUCTION} 

Although a robust paradigm for star formation within giant molecular
cloud complexes now exists (Shu et al. 1987), most stars are thought
to be born within some type of cluster environment (e.g., Lada \& Lada
2003; Porras et al. 2003). Furthermore, the distribution of cluster
sizes (given here in terms of stellar membership $N$) remains
uncertain, and the influence of the cluster environment on forming
stars and planets is not completely understood. One important way in
which the cluster setting can influence the formation of additional
cluster members, and especially their accompanying planetary systems,
is through the radiation fields provided by the background
environment.  This radiation can potentially drive a number of
significant processes, including [1] the heating of starless cores,
leading to evaporation and the loss of star forming potential (e.g.,
Gorti \& Hollenbach 2002), [2] the evaporation of circumstellar disks,
leading to loss of planet forming potential (e.g., Shu et al.  1993,
Hollenbach et al. 1994, St{\"o}rzer \& Hollenbach 1999, Adams et
al. 2004), [3] ionization within starless cores, leading to greater
coupling between the magnetic fields and gas (e.g., Shu 1992), thereby
acting to suppress continued star formation, and [4] ionization of
circumstellar disks, which helps maintain the magneto-rotational
instability (MRI), which in turn helps drive disk accretion (e.g.,
Balbus \& Hawley 1991). For applications to circumstellar disks, note
that the background radiation from the cluster environment often
dominates that produced by the central star (e.g., Hollenbach et
al. 2000; Adams \& Myers 2001); this claim is substantiated and
quantified by the results of this paper.

The goal of this paper is to provide a systematic construction of the
distributions of the expected radiation fields in clusters, including
UV luminosities and fluxes, in both the EUV and FUV radiation bands
(these are defined in \S 2).  With these distributions of luminosities
and fluxes determined, one can then assess the importance of the
cluster background environment for star and planet formation, through
the channels outlined above (and others).  Although a complete
assessment of this type has not been done, the importance of radiation
fields in clusters has been emphasized in previous work. A study of
the EUV radiation fields and their potentially harmful effects on
planet formation has been carried out (Armitage 2000).  An analogous
treatment of the FUV radiation fields for the clusters in the solar
neighborhood has also been done (Adams et al. 2006; hereafter
APFM). Finally, Parravano et al. (2003) have reconstructed the UV
radiation fields in the local interstellar medium, where the ultimate
source of the UV radiation is star forming regions.

This paper is organized as follows. In \S 2 we outline the basic
approach, including specification of the cluster sample, assumptions
about the gas content, and our characterization of the stellar IMF.
We then present the distributions of EUV and FUV luminosities in \S 3,
where these distributions are constructed both for clusters of a given
size $N$ and for the entire ensemble of cluster sizes. In \S 4, we
present the corresponding distributions of EUV and FUV flux, which
requires the additional specification of the distribution of radial
positions within the cluster.  This section also discusses the effects
of extinction and the effects of averaging over stellar orbits on the
resulting distributions. We conclude in \S 5 with a summary of our
findings and a discussion of potential applications.

\section{FORMULATION} 

In this section, we specify the input parameters used to produce the
resulting distributions of radiation fields. We must specify the
distribution of cluster membership sizes $N$, the cluster radii as a
function of $N$, the stellar initial mass function (IMF), and the
mass-luminosity relationship for massive stars. We defer a discussion
of the gas content of clusters, and our treatment of extinction, until
\S 4 where we consider the flux distributions.  Throughout this paper,
we often present results as a function of cluster size $N$. In this
context, we consider the cluster to have a stellar membership of $N$
primaries and we ignore binarity. We also present results for both FUV
radiation, which is defined to have photon energies in the range 6 eV
$\le h \nu \le$ 13.6 eV, and EUV radiation, where $h \nu \ge$ 13.6 eV.
To define nomenclature: We use the general subscript `$UV$' to denote
either of these ultraviolet radiation bands, and the explicit
designation `$EUV$' or `$FUV$' to denote one of the two particular UV
radiation bands.

\begin{figure}
\figurenum{1}
{\centerline{\epsscale{0.90} \plotone{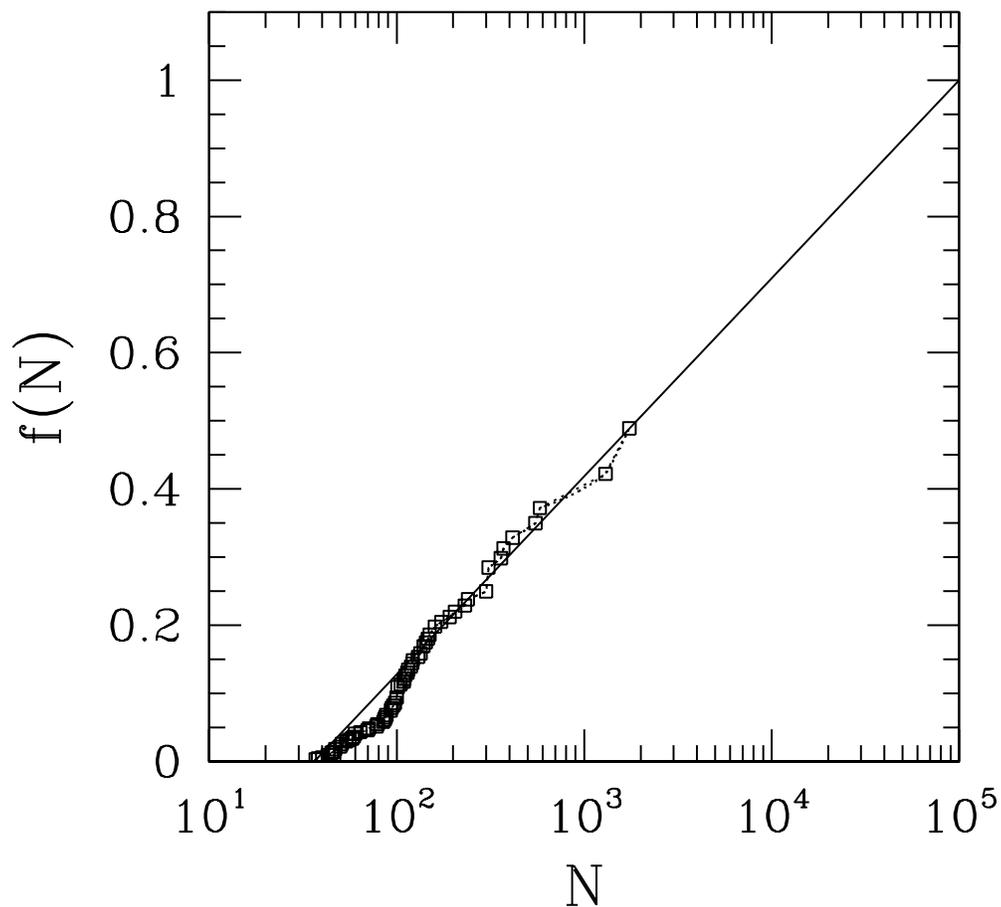} }}
\figcaption{Cumulative distribution showing the fraction of stars born
in clusters of stellar membership $N$ as a function of $N$. The data
points with the dotted curve show the observed distribution in the
solar neighborhood (as compiled in Lada \& Lada 2003). The solid
line shows an extended distribution for which the same trend --
roughly equal number of stars born in each decade of cluster size --
extends up to $N = 10^5$. Note that the scale for $f(N)$ is given 
for the extended distribution; for the solar neighborhood, $f(N)$ = 1 
for $N$ = 1740. } 
\label{fig:clustnum} 
\end{figure}

For the distribution of cluster sizes (stellar membership $N$), we use
two standard choices. The first is the observed distribution of
cluster sizes in the stellar neighborhood.  In this case, both Lada \&
Lada (2003) and Porras et al. (2003) have produced distributions
of observed clusters out to 2 kpc (1 kpc) and down to cluster sizes of 35
(30), respectively. These observational compilations show that the
number of stars born in clusters of size $N$ is evenly distributed
logarithmically over the range of clusters seen in the solar
neighborhood, i.e., from $N$ = 30 to 2000. For the sake of
definiteness, we use the actual data from the Lada \& Lada (2003)
compilation as our first standard distribution.  While this 
catalog is not complete, it is likely representative of the
basic statistical properties of embedded clusters within 2 kpc.
For our second
standard distribution, we assume that this general trend continues up
to larger cluster sizes, i.e., that star formation takes place in
clusters with an even logarithmic distribution extending up to $N$ =
$10^5$ stars. Both the Lada/Lada distribution and the extended
distribution are shown in Figure \ref{fig:clustnum}.

The radial sizes of observed clusters follow a well-defined law of 
the form 
\be
\rclust (N) = R_0 \left( {N \over 300} \right)^{1/2} \, , 
\label{rofn}
\ee 
where the scale $R_0 \approx 1$ pc (see Fig. 2 of APFM,
which uses the data from Lada \& Lada 2003, and Carpenter 2000).  
We use this empirically determined law to specify cluster radii
throughout this paper, including for the extrapolation of the cluster
distribution described above.

The stellar IMF has a
power-law form, with a nearly universal slope, for masses $M_\ast 
\simgreat 1 M_\odot$ (starting with Salpeter 1955) and a lognormal 
form below this value.  
Since the total luminosity of a star scales roughly as $m^3$ 
for stars with mass less than $\sim 10 M_\odot$ and
the IMF has a slope of $\sim 2.35$, the luminosity distribution
as a function of stellar mass scales as $\sim m^{0.7}$ for 
intermediate mass stars.  Of course, low mass stars
fall within the lognormal part of the IMF and 
have spectra that peak well below the UV band.   As a
result, we can ignore the contribution of all stars smaller than 1
$M_\odot$ to a reasonable approximation (as quantified in Fig. 2 below).  
To specify the
initial mass function in this context, we only need to correctly
account for the fraction $\fone$ of stars with $M_\ast > 1 M_\odot$
and the slope $\Gamma$ at high stellar masses. We thus assume that the
stellar IMF has a power-law form for mass $M_\ast > 1 M_\odot$ with
index $\Gamma$, i.e.,
\be
{dN_\star \over dm} = A m^{-\Gamma} \, = \fone (\Gamma - 1) m^{-\Gamma} \, , 
\ee
where $m$ is the mass in units of solar masses and where the slope
$\Gamma = 2.35$ for the classic form of Salpeter (1955). Although this
slope is remarkably consistent over a wide variety of regions (Massey
2003 and references therein), we can account for possible variations
in the IMF by allowing the index $\Gamma$ to vary. In the second
equality, we have normalized the distribution according to the
convention
\be
\int_1^{m_{max}} {d N_\star \over d m} dm = \fone \, , 
\label{normalize} 
\ee
so that $\fone$ is defined to be the fraction of the stellar
population with masses larger than 1 $M_\odot$.  For a typical stellar
mass function (e.g., that advocated by Adams \& Fatuzzo 1996), the
fraction $\fone \approx 0.12$. Notice also that we assume that the IMF
does not extend up to arbitrarily high stellar masses, but rather is
truncated at a maximum mass $m_{max}$. In this context, the IMF is
thus determined by the parameter set $(\fone, \Gamma, m_{max})$. In
this paper, we fix $\fone$ = 0.12 for all cases and vary the other two
parameters such that $(\Gamma, m_{max})$ = (2.35, 100), (2.1, 100),
(2.5, 100), and (2.35, 120). We adopt the first of these choices of
IMF parameters as our standard case, but explore the effects of
varying the IMF in \S 4.

\begin{figure}
\figurenum{2}
{\centerline{\epsscale{0.90} \plotone{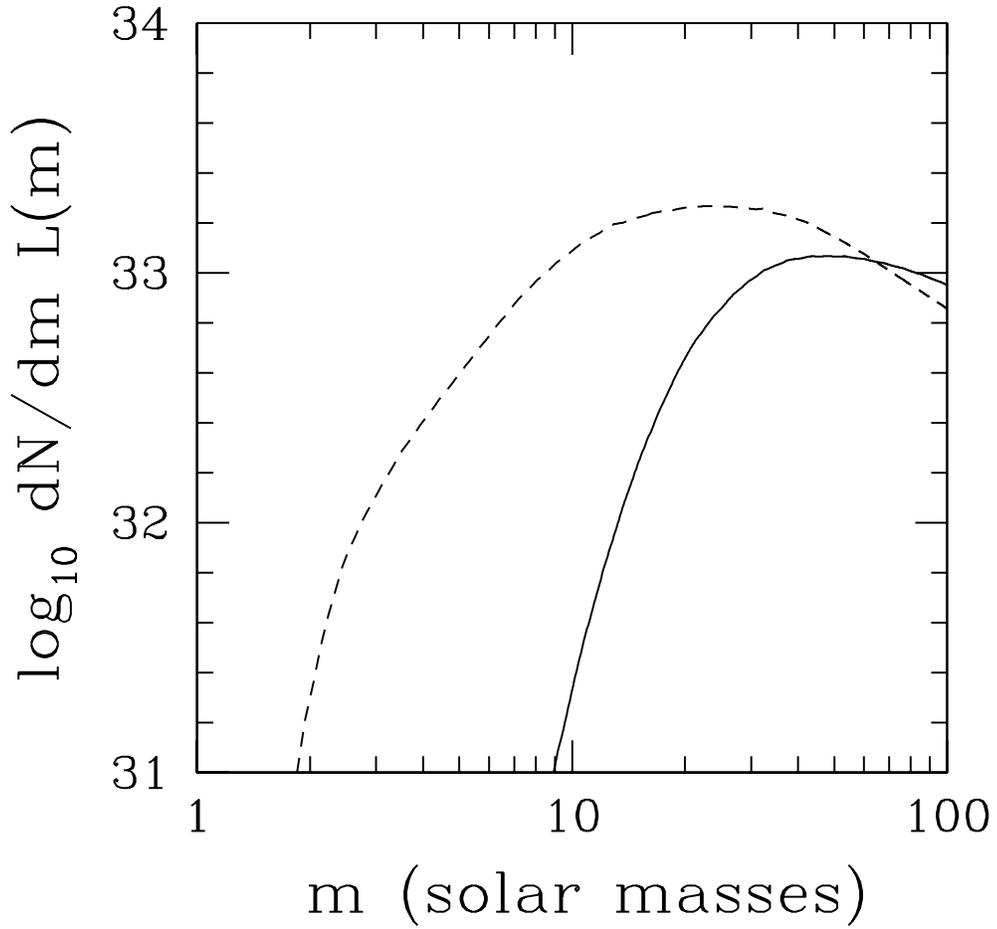} }}
\figcaption{Distribution of UV luminosity over the range of stellar
masses.  The solid line shows relative amount of EUV luminosity
produced by stars of varying mass, plotted as a function of mass.  
The dashed line shows the analogous distribution for FUV luminosity. 
The vertical axis gives the distribution in units of erg s$^{-1}$ 
$M_\odot^{-1}$. The IMF is normalized as in equation (\ref{normalize}). } 
\label{fig:uvmass} 
\end{figure}

With the IMF specified, we need to determine the relationship between
stellar mass and UV luminosity. Toward that end, we use stellar models
calculated by Maeder and collaborators (e.g., Maeder \& Meynet 1987;
Schaller et al. 1992). Specifically, this previous work provides a
grid of stellar models as a function of both mass and age. In this
setting we use the zero age models to specify the stellar luminosity
and effective temperature. We then follow the procedure of Armitage
(2000) to determine the luminosity in the EUV and FUV bands (note that
departures from blackbody spectra are important in this regime).
Notice also that the UV radiation is dominated by the largest stars,
which reach the main-sequence rapidly, so zero-age main sequence
models are appropriate for the construction of these luminosity
distributions (i.e., we need not consider pre-main-sequence evolution).
Additional stellar evolution results in two competing effects for the
total cluster luminosity: All stars get brighter as they age, but the
most massive stars (with $m \sim 100$) have relatively short lifetimes.
Adopting zero-age main sequence values for the luminosity, as done 
here, thus yields one representative outcome for a given cluster. 
The next level of complication, not done here, would be to construct 
distributions of UV luminosity and flux as a function of cluster age. 

Figure \ref{fig:uvmass} shows the range of stellar masses that
dominates the production of UV radiation for both the EUV and FUV
bands. The dashed curve shows the standard IMF multiplied by the FUV
luminosity, plotted as a function of stellar mass. The distribution
shows a broad peak near 20 $M_\odot$. This result vindicates our
assumption that low mass stars (defined here to be those with masses
$M_\ast < 1 M_\odot$) have a negligible contribution to the total UV
luminosity of the cluster. Similarly, the solid curve in Figure
\ref{fig:uvmass} shows the stellar IMF multiplied by the EUV
luminosity as a function of mass. This distributed is skewed toward
even larger stellar masses, with a peak near 40 $M_\odot$.

For a given stellar IMF, we thus obtain corresponding distributions of
FUV and EUV luminosities. To leading order, these distributions can be
characterized by their expectation values and their width (or variance);
at a higher order of analysis, however, the shapes of the distribution
show significant departures from simple gaussians. For a given IMF and a
given UV band (either FUV or EUV), the expectation value $\uvexp$ of the
UV luminosity is determined by the integral
\be
\uvexp = \int_1^\infty L_{UV}(m) {dN_\star \over dm} dm \, , 
\label{meanuv} 
\ee 
where $dN_\ast/dm$ = 0 for stellar masses $m > m_{max}$.  The
expectation value of equation (\ref{meanuv}) is normalized so that it
provides the expected UV luminosity {\it per star}. Because of the
wide range of stellar masses and the sensitive dependence of both EUV
and FUV emission on stellar mass, this expectation value is much
larger than the UV radiation emitted by the majority of stars. As a
result, the UV radiation from a cluster will converge to the value
implied by this expectation value only in the limit of large $N$
(the value of $N$ necessary to be ``large'' is determined below). 
Furthermore, since the EUV radiation depends even more sensitively on
stellar mass, this trend is more extreme for the case of EUV radiation
(compared to FUV radiation). Small clusters will generally display
large departures from the expectation value.

Here we need to determine both the expectation value and variance of
the UV luminosity distribution.  The UV luminosity is that of the
entire cluster, and is given by the sum
\be
L_{UV} (N) = \sum_{j=1}^N L_{UVj} \, , 
\label{uvsum} 
\ee
where $L_{UVj}$ is the UV luminosity from the $jth$ member.  In this
formulation, we assume that the UV luminosity for a given star is
determined solely by the stellar mass, which is drawn independently 
from a specified stellar IMF. This sum is thus the sum of random
variables, where the variables (the individual contributions to the UV
power) are drawn from a known distribution, which is in turn
determined by the IMF and the $L_{UV}-m$ relation.  In the limit of
large $N$, the expectation value of the UV power is given by
\be L_{UV} (N) = N \uvexp \, .
\label{lnexp} 
\ee 
As usual, the central limit theorem implies that the distribution of
values $L_{UV} (N)$ must approach a gaussian form as $N \to \infty$
(e.g., Richtmyer 1978), although convergence is often slow. One of the
issues of interest here is the value of stellar membership $N$
required for these statistical considerations to be valid; similarly,
we would like to know the fraction of the cluster population that has
such sufficiently large $N$. In its limit of applicability, this 
gaussian form for the composite distribution is independent of the
form of the initial distributions, i.e., it is independent of the
stellar IMF and the mass-luminosity relation. The width of the
distribution also converges to a known value given by the expression
\be 
\sigbar^2 = {1 \over N}
\sum_{j=1}^N \sigma_j^2 \quad \Rightarrow \quad \sigbar = \sqrt{N}
\sigma_0 \, , 
\label{nvariance} 
\ee 
where $\sigma_0$ is the width of the individual distribution, i.e., 
\be 
\sigma_0^2 \equiv \langle L_{UV}^2 \rangle - \uvexp^2 \, .  
\ee 

The expectation values and widths of the luminosity distributions are
listed in Table 1 for the four types of stellar IMF used in this
paper.  The first column gives the parameters $(\Gamma, m_{max})$ of
the IMF, where the fraction $\fone$ = 0.12 for all of the cases.  The
expectation values are listed as the UV luminosity (either EUV or FUV)
per star, and are presented in units of erg/s (cgs units).  The widths
of the distributions are given in normalized form, where ${\widetilde
\sigma}_{0UV} \equiv \sigma_{0UV}/\langle L_{UV} \rangle_\ast$.
Although the luminosity distributions, as characterized by their width
and expectation values, vary somewhat with the stellar IMF, they all
show the same general features. For all four IMFs and both UV bands,
the expectation values (per star) are $\uvexp \sim 10^{36}$ erg/s in
order of magnitude, or about 250 $L_\odot$.  In other words, the mean
UV luminosity is always much larger than the total luminosity of the
typical star (which has mass $M_\ast \sim 0.5 M_\odot$ and luminosity
$\sim0.1 L_\odot$ on the main-sequence). This finding is simply a
manifestation of the sensitive dependence of the UV luminosity on
stellar mass and the large UV luminosities produced by the most
massive stars. The second feature illustrated in Table 1 is that the
widths of the luminosity distributions are large, roughly 20 -- 40
times wider than the expectation values. This property of the
distributions implies that one needs a large number of cluster stars
$N$ in order to fully sample the distribution. For example, in order
for the width of the distribution to be narrower than the expectation
value, the number of stars $N$ must satisfy the relation
\be 
N > N_{min} = {\widetilde \sigma}_{0UV}^2 \sim 10^3 \, , 
\ee
where the numerical value depends on the assumed IMF, but is of order
1000 (see also APFM). Notice that 90\% of the stars in the observed
sample in the solar neighborhood are born in clusters with $N < 1000$,
i.e., the regime of incomplete sampling of the IMF.  For the
extrapolated distribution of cluster sizes, only about half of the
stars are born in the regime of incomplete sampling (see Fig.
\ref{fig:clustnum}). In either case, a significant fraction of stars
are born in clusters where the UV luminosity is subject to incomplete
sampling. This finding implies that one must consider the full
distribution of possible luminosities --- not just mean or median
values --- when assessing the importance of radiation fields in these
environments.

\begin{table}
\begin{center}
\caption{\bf Parameters for UV Luminosity Distributions} 
\medskip 
\begin{tabular}{lcccc} 
\hline 
\hline 
IMF $(\Gamma, m_{max})$ 
& $\langle L_{FUV} \rangle_\ast$ (erg/s) & ${\widetilde \sigma}_{0FUV}$ &  
$\langle L_{EUV} \rangle_\ast$ (erg/s) & ${\widetilde \sigma}_{0FUV}$ \\ 
\hline
(2.35, 100)  & 1.23 $\times 10^{36}$ & 25.5  & 8.47 $\times 10^{35}$ & 36.9 \\
(2.1, 100)   & 2.53 $\times 10^{36}$ & 18.7  & 1.89 $\times 10^{36}$ & 25.5 \\
(2.5, 100)   & 7.91 $\times 10^{35}$ & 30.4  & 5.15 $\times 10^{35}$ & 46.2 \\
(2.35, 120)  & 1.36 $\times 10^{36}$ & 26.6  & 1.01 $\times 10^{36}$ & 38.5 \\
\hline 
\hline 
\hline 
\end{tabular}
\end{center} 
\end{table} 

\section{ULTRAVIOLET LUMINOSITY DISTRIBUTIONS} 

In this section, we construct the distributions of ultraviolet
luminosities. This determination is done for both FUV and EUV
radiation, and is carried out in two ways.  First, we construct the UV
luminosity distribution for clusters of a given size $N$. In this
case, each cluster independently samples the stellar IMF $N$ times and
the resulting sampling variation leads to a distribution of possible
UV luminosities.  Second, we construct the UV luminosity distribution
for a collection of clusters over a distribution of cluster sizes
(namely those shown in Fig. \ref{fig:clustnum}). In this latter case,
the distribution of UV luminosity (both EUV and FUV) is determined by
two input distributions, the stellar IMF and the distribution of
cluster membership $f(N)$.

\begin{figure}
\figurenum{3}
{\centerline{\epsscale{0.90} \plotone{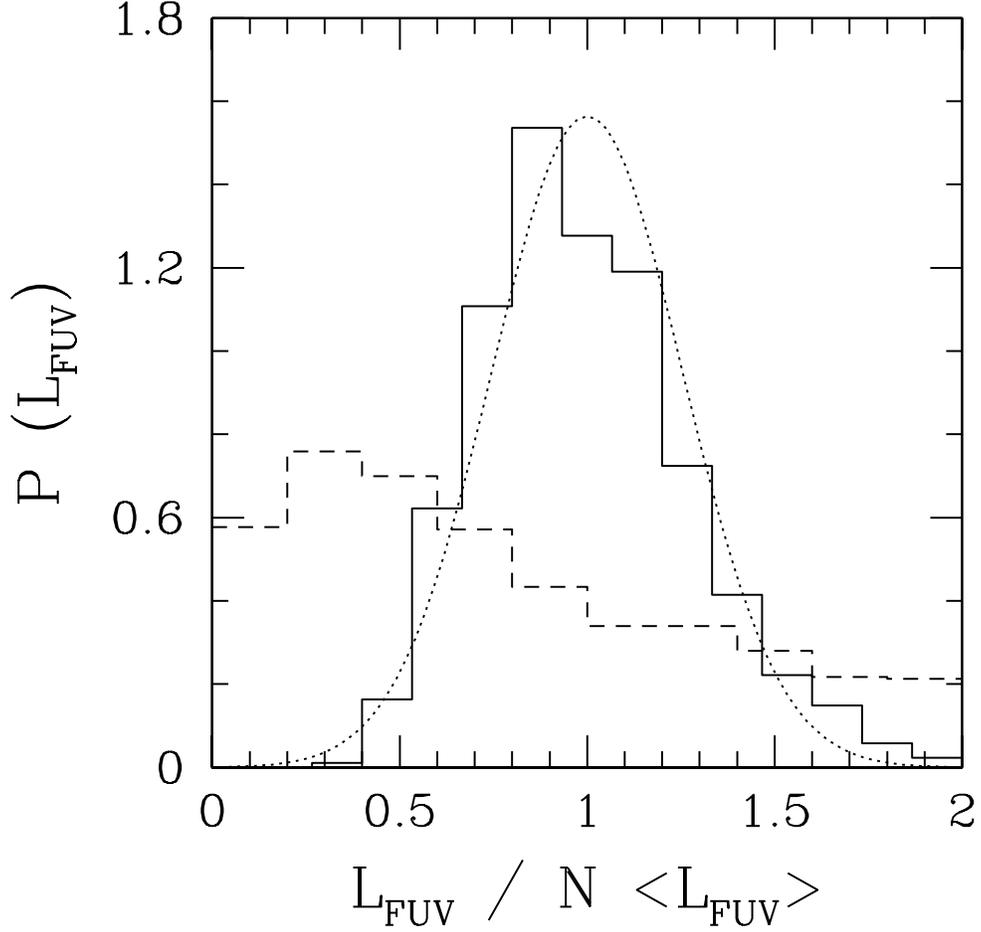} }}
\figcaption{Distribution of FUV luminosities. The dashed histogram
shows the distribution for clusters with $N$ = 1000 stars; the solid
histogram shows the distribution for $N = 10^4$. Note that the
luminosities (along the horizontal axis) are normalized by the average
value one would get with complete sampling of the stellar IMF.  The  
probability distribution (vertical axis) is normalized so that the 
area under the curves (the total probability) is unity. For
comparison, the dotted curve shows a gaussian distribution with width
$\sigma = 25.5 / \sqrt{10^4} \approx$ 0.255 (see Table 1). } 
\label{fig:gauss} 
\end{figure}

Before constructing the physically relevant luminosity distributions,
we first consider the shapes of the distributions, in particular their
substantial departures from gaussianity.  To illustrate this behavior,
Figure \ref{fig:gauss} presents the distribution of FUV
luminosities, normalized so that the area under the curves equals
unity.  This plot uses the FUV luminosity divided by the
expected mean FUV luminosity for the horizontal axis, and thus
emphasizes the {\it shape} of the distributions. Keep in mind that the
dimensionless widths of the luminosity distributions discussed above
(see Table 1) apply to the individual distributions (per star),
whereas the distributions in Figure \ref{fig:gauss} give the shape for
different values of $N$.  For relatively ``small'' $N$, i.e., $N$ =
1000 as shown by the dashed histogram, the distribution is distinctly
non-gaussian and peaks at a value significantly less then the mean.
For larger $N$, however, the distribution attains the expected
gaussian form, as shown by the solid histogram for $N = 10^4$. The
dotted curve shows the corresponding gaussian distribution with the
expected dimensionless width for the $N$ = $10^4$ case; notice the
good agreement.  This plot shows that clusters must have relatively
large stellar membership, $N$ as large as $N \sim 10^4$, before the
distribution of FUV luminosity is not dominated by sampling
statistics. Note that no clusters this large are observed within 2 kpc
(Lada \& Lada 2003), so that the entire solar neighborhood is subject
to these sampling variations.

\begin{figure}
\figurenum{4a}
{\centerline{\epsscale{0.90} \plotone{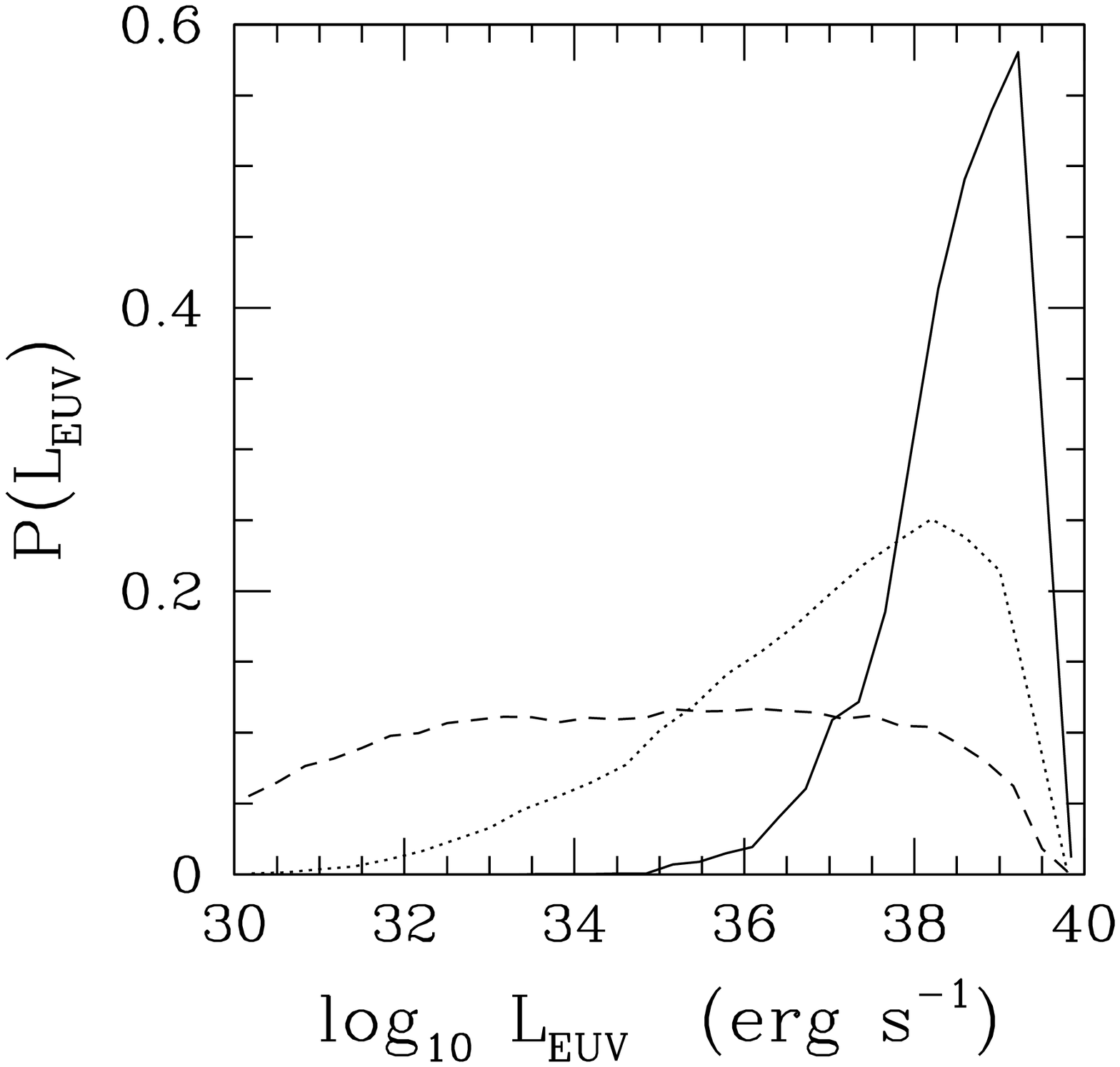} }}
\figcaption{}  
\label{fig:lumin_euv} 
\end{figure}

Figure 4 shows the probability distribution for EUV luminosities
(Fig. 4a) and FUV luminosities (Fig. 4b) for three different cluster 
sizes $N$.  Note that these distributions (along with 
those shown in Figures 6 - 9 below) are built using the base 10 logarithm 
of the luminosity (flux).  Specifically, in Figs. 4 and 6 -- 9, $P(x)$ d(log$_{10}x$)
(where $x$ is either $L_{UV}$ or $F_{UV}$) is the probability that $x$ falls between
log$_{10}x$ and log$_{10}x$+d(log$_{10}x$). 
Note that the mean of the logarithm is not the same as the logarithm
of the mean, $\langle \log_{10} x \rangle \ne \log_{10}\langle x
\rangle$,
so that even a gaussian distribution is skewed (the peak of the
distribution
falls to the right of mean) when plotted as in Figure 4.

In both Figures 4a and 4b, distributions are shown for $N$ = 100
(dashed curves), $N$ = 300 (dotted curves), and $N$ = 1000 (solid
curves). As the number of cluster members $N$ increases, the
distributions shift to the right, toward higher luminosities, as
expected. The distributions also appear to become narrower with
increasing $N$.  The relative width $\sigbar/N \uvexp$ does indeed
become smaller as $N$ grows larger, as outlined above, but the
apparent decrease in the total width with $N$ is an artifact of
plotting the distribution using a logarithmic scale for the UV
luminosity (on the horizontal axis).

\begin{figure}
\figurenum{4a}
{\centerline{\epsscale{0.90} \plotone{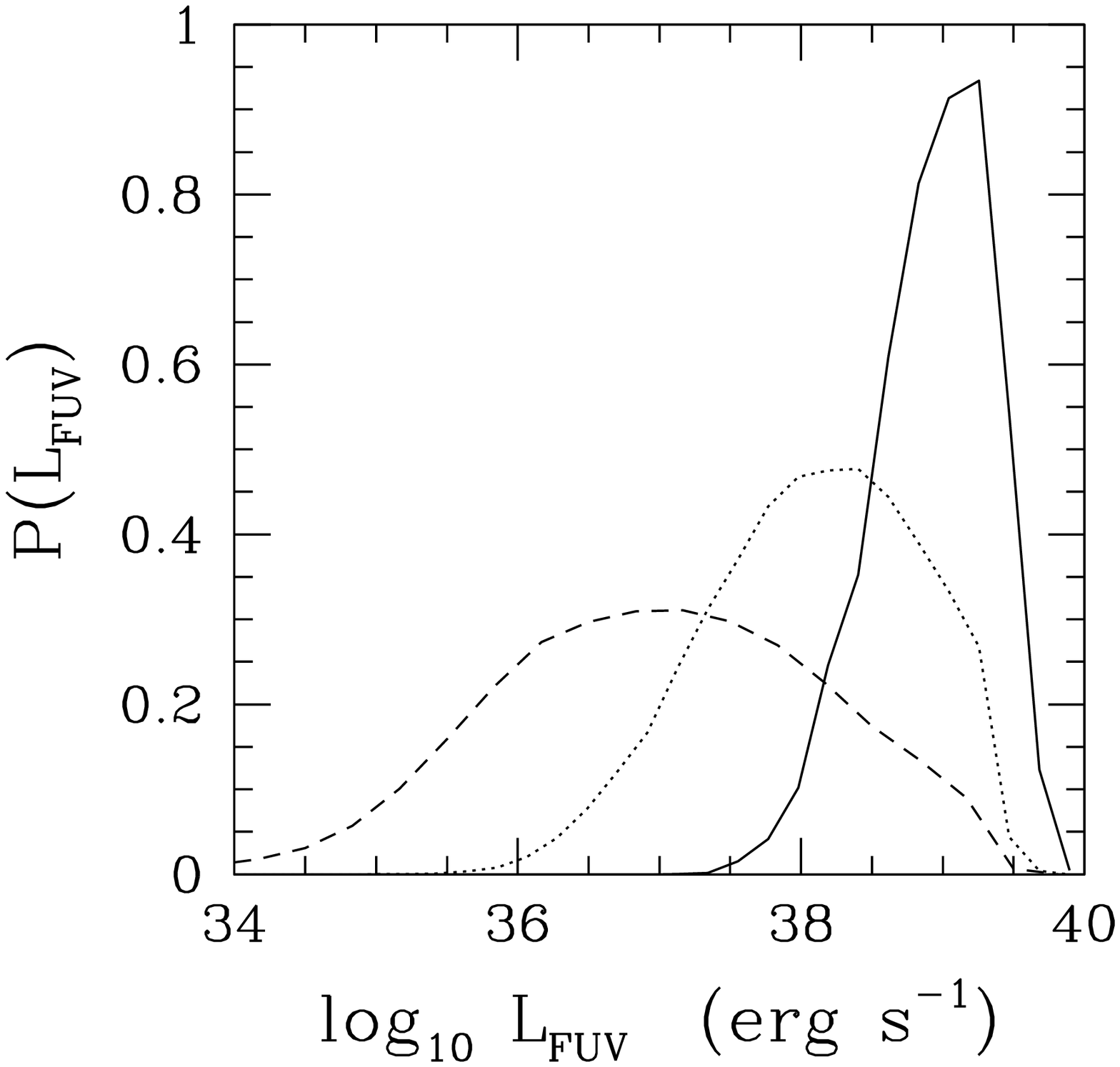} }}
\figcaption{Probability distribution for EUV luminosities. The three 
curves show the normalized distributions for clusters of varying
sizes, with $N$ = 100 (dashed), $N$ = 300 (dotted), and $N$ = 1000
(solid). Each probability distribution is built up from $10^7$
independent realizations of the cluster population. 
b. -- Same as Figure 4a, but for FUV luminosities. }  
\label{fig:lumin_fuv} 
\end{figure}

Figure 5 shows the UV luminosities for clusters as a function of
cluster membership size $N$.  Results are shown here for both the EUV
luminosities (Fig. 5a) and the FUV luminosities (Fig. 5b).  In each
case, the luminosity for a given cluster size $N$ shows a wide range
of values due to incomplete sampling of the IMF, i.e., the UV
luminosity for a given $N$ is characterized by a distribution.  The
mean value (for a fully sampled IMF) follows the solid line shown in the
Figures. The median value, which is significantly smaller than the
mean (for a fully sampled IMF)
at small $N$, is shown by the data point symbols. Here the width
of the distribution is delineated in two different ways. The error
bars show the range of luminosities enclosing the fraction of the
distribution from 16.5\% to 83.5\% of the total.  The dotted curves
show the total expected range of luminosities calculated from the
statistical considerations above, i.e., the expectation value plus or
minus the width of the distribution for a fully sampled IMF.  Since the luminosity
distribution is wide, and far from gaussian at low values of $N$, the
width of the distribution defined this way is larger than the
expectation value. This property of the distributions results in the
lower dotted line falling rapidly toward zero (becoming nearly
vertical) at $N \approx 1200$ for the EUV distribution and at $N
\approx 700$ for the FUV distribution. These results are in basic
agreement with those obtained earlier for EUV radiation (Armitage
2000) and FUV radiation (APFM).

The results depicted in Figure 5 show that the distributions of UV
luminosity have qualitatively different behavior for large $N$ and
small $N$ clusters. Note that the ranges of expected luminosity values
become centered on the expectation values for sufficiently large
values of stellar membership $N$.  For the EUV distribution, this
centering occurs for $N \simgreat 2000$, whereas for the FUV
distribution it occurs for $N \simgreat 1000$. For both cases, this
centering occurs at somewhat larger $N$ than the values required for
the distributions to be narrower than their expectations values.  By
coincidence, the required ``centering'' values are roughly the same as
the value of $N$ for the largest cluster (the Orion Nebula Cluster, or
ONC) in the solar neighborhood cluster distribution. In other words,
for all of the clusters in the solar neighborhood, the stellar
membership is {\it not} large enough to use statistical arguments to
predict expectation values, etc., so that the central limit theorem
does not fully apply (i.e., the limit of large $N$ is not reached).
The distributions thus depend sensitively on the sampling of the
underlying IMF.  Notice that this finding makes sense: For $N \sim
1000$, say, the number $N_{UV}$ of stars large enough to provide {\it
any} significant UV luminosity is only about $N_{1} \sim \fone N \sim
120$, whereas the number $N_{20}$ of stars larger than 20 $M_\odot$
(where most of the UV is emitted --- see Fig. \ref{fig:uvmass}) is
only $N_{20} \sim N \fone (20)^{-1.35} \sim 2$. In other words, in
rough terms, only clusters with stellar membership $N$ greater than
$\sim$1000 are large enough to populate the part of the IMF where most
of the UV radiation is emitted.

\begin{figure}
\figurenum{5a}
{\centerline{\epsscale{0.90} \plotone{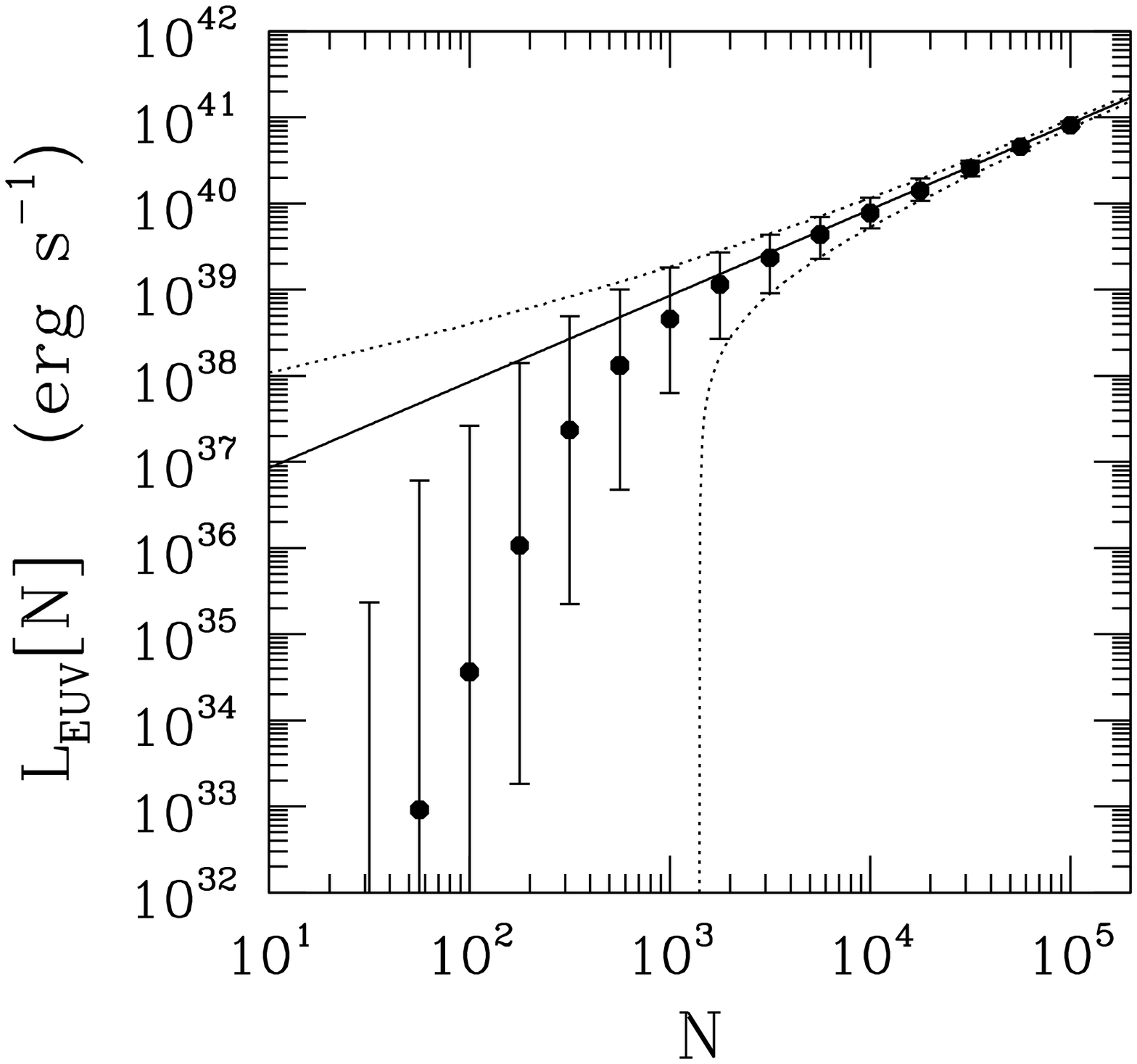} }}
\figcaption{} 
\label{fig:lumnumber_euv} 
\end{figure}

\begin{figure}
\figurenum{5a}
{\centerline{\epsscale{0.90} \plotone{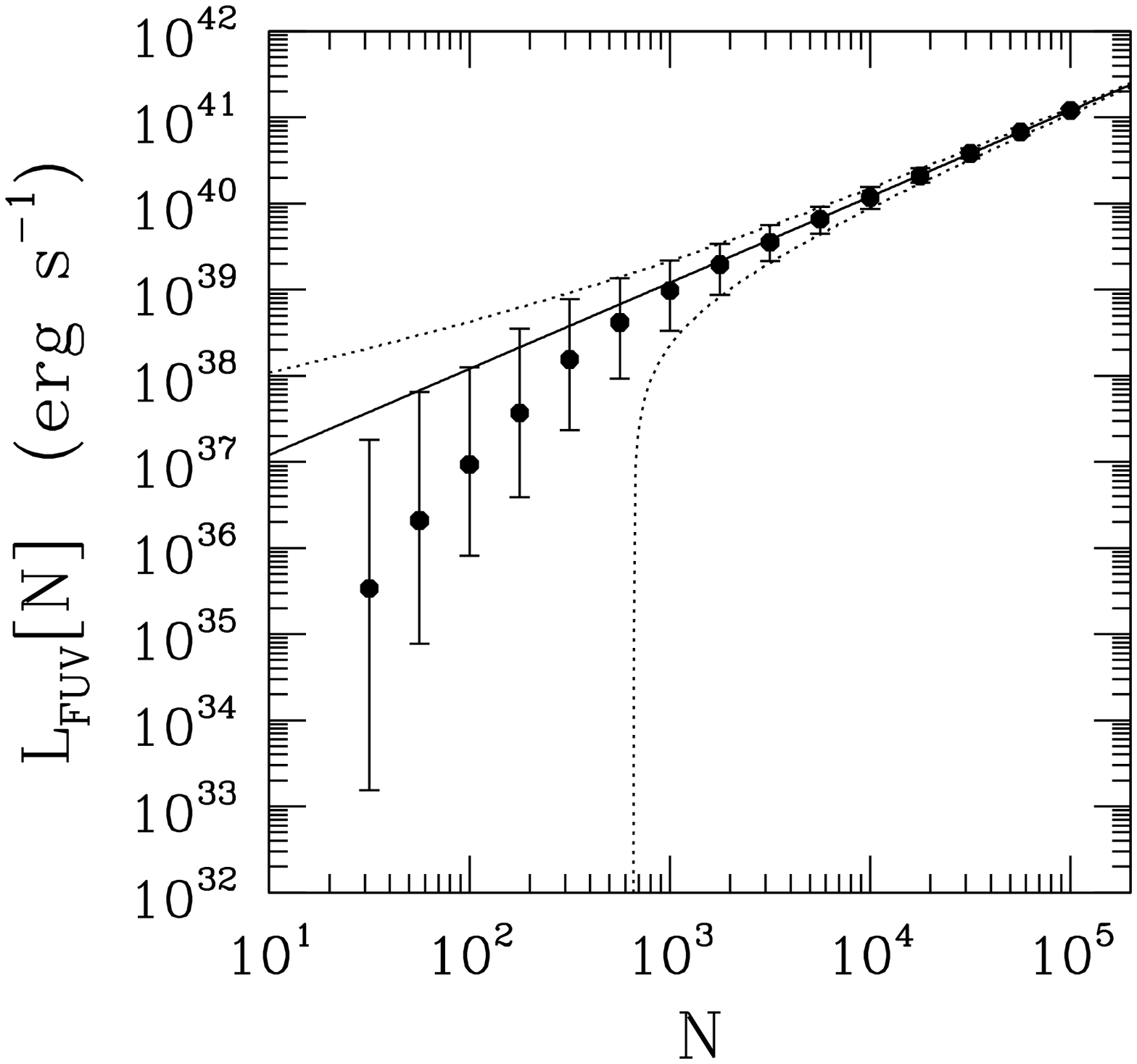} }}
\figcaption{EUV Luminosity per cluster as a function of cluster size
$N$.  For each value of $N$, the EUV luminosity can have a wide
distribution of values.  The solid line shows the mean value
for a fully sampled IMF ($N \uvexp$) as a
function of cluster membership $N$.  
The data points show the median
values of the distribution, where the error bars delineate the range
from 0.165 to 0.835. The dotted curves show the expectation value 
plus or minus the width of the distribution for a fully sampled IMF. 
b. -- Same as Figure 5a, but for the FUV Luminosity per cluster.} 
\label{fig:lumnumber_fuv} 
\end{figure}

The distributions presented thus far have been constructed using the
observed distributions of cluster sizes $N$ in the solar neighborhood
(Lada \& Lada 2003). We now consider the effects of variations in the
cluster size distribution on the luminosity distributions.
Specifically, we first sample the cluster size distribution to
determine the size $N$ of the cluster to which a ``test'' star
belongs, and then sample the IMF N times to determine the
corresponding luminosity of our ``test'' star's cluster.  We repeat
this process a total of $10^7$ times to then build up luminosity
distributions (we have varied the sampling size and verified that it
is large enough for convergence).  Figure 6 shows a comparison of the
resulting luminosity distributions for different cluster size
distributions, including the solar neighborhood (solid curve) and the
extended distribution that is extrapolated up to cluster sizes of $N =
10^5$ (dashed curves). The EUV distributions are shown in Figure 6a,
and the FUV distributions are shown in Figure 6b. A related issue is
the extent to which the UV radiation fields are dominated by the most
massive star in the cluster. To consider this problem, both panels of
Figure 6 also show the distributions of luminosity using only the most
massive star in the cluster (dotted curves), for both the solar
neighborhood and extended distributions of cluster sizes.  For the
solar neighborhood distribution, which includes only relatively
``small'' clusters, the distribution of luminosity with only the most
massive star and the distribution with all stars are nearly the same;
this result indicates that the UV radiation fields are dominated by
the single most massive star in such systems. In the case of the
extended cluster size distribution, however, there is marked
difference between the distribution that includes all stars and the
one that includes only the most massive star. This trend is clearly
evident for both the EUV and FUV bands, and shows that many stars
provide a significant contribution to the UV luminosity for large
clusters.
 
\begin{figure}
\figurenum{6a}
{\centerline{\epsscale{0.90} \plotone{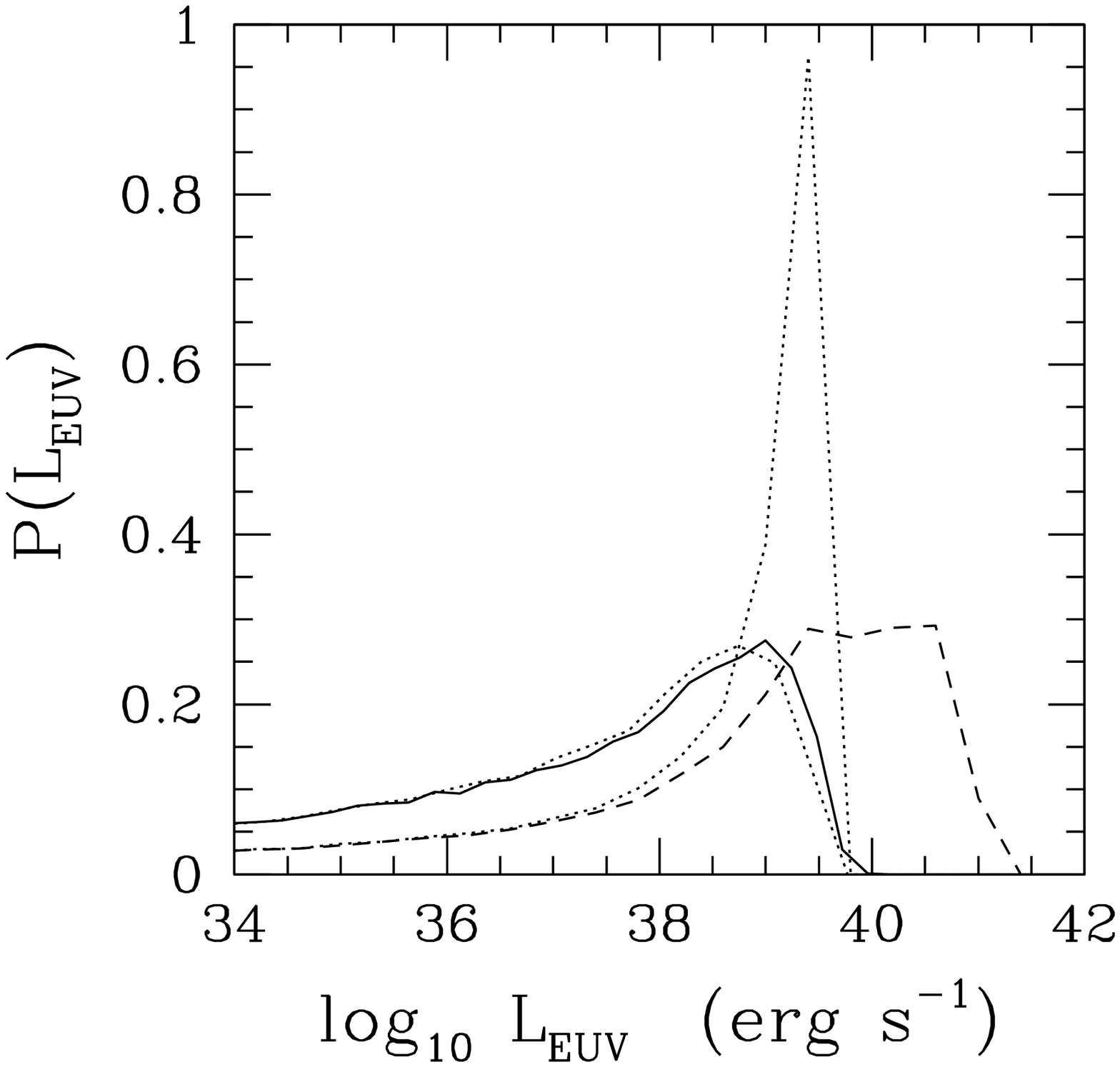} }}
\figcaption{}
\label{fig:lumcluster_euv} 
\end{figure}

\begin{figure}
\figurenum{6a}
{\centerline{\epsscale{0.90} \plotone{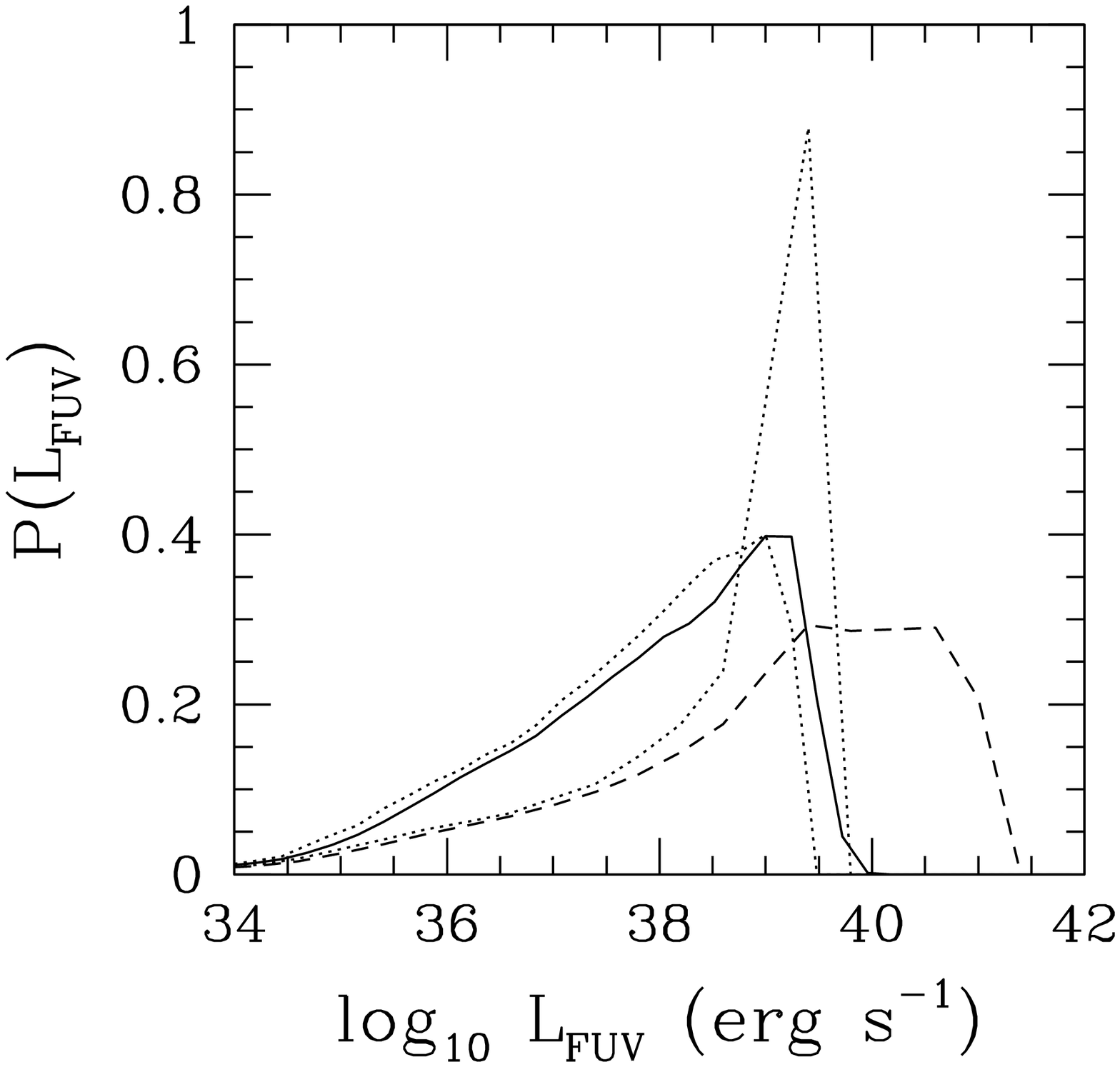} }}
\figcaption{Distributions of EUV luminosities for different cluster
samples. The solid line shows the distribution of EUV luminosity for
the sample of clusters observed in the solar neighborhood (Lada \&
Lada 2003). The dashed curve shows the corresponding distribution of
EUV luminosities for the extended cluster distribution that extends up
to $N$ = $10^5$. The dotted curves show the analogous distributions
for the case in which only the most massive star is allowed to
contribute to the EUV luminosity of the cluster. 
b. -- Same as Figure 6a, but for FUV luminosities.
}  
\label{fig:lumcluster_fuv} 
\end{figure}

\section{ULTRAVIOLET FLUX DISTRIBUTIONS} 

Given the distribution of UV luminosity, both for clusters of a given
size $N$ and for the expected distribution of clusters, the next step
is to construct distributions of UV flux. To convert the distributions
of EUV and FUV luminosities into fluxes, we first need to specify the
distribution of stars within clusters.  In other words, we must
convolve the distribution of luminosities with a distribution of
radial positions to produce a distribution of fluxes.  For most of
this paper, we assume that the stars follow a distribution of radial
positions that corresponds to a density law of the form $\rho_\ast
\propto 1/r$, or, equivalently, $dm \propto r^2 \rho_\ast dr \propto r
dr$. Although the density profile will have an inner cutoff (to remove
the apparent singularity at the origin), this scale does not enter
into the problem (one can integrate over it). However, the outer
boundary does play a role and here we truncate the distribution at the
outer radius $\rclust$ of the cluster, where $\rclust$ is given as a
function of $N$ in equation (\ref{rofn}). This form for the density
distribution of stars is consistent with results from N-body
simulations of analogous clusters (see eq. [3] and Table 2 of APFM).
Note that by considering the distribution of radial positions to
determine flux, we are determining the complete ensemble of possible
flux values provided by the cluster environment. However, individual
solar systems will execute orbits within the cluster potential, and
the distribution of orbit-averaged fluxes will be somewhat
narrower. We take up this issue in \S 4.4.

\subsection{Extinction} 

In this treatment, we provide flux distributions both with and without
extinction. Since the gas (and hence the dust) in young clusters has a
relatively short lifetime, perhaps only $\sim3$ Myr (Allen et
al. 2007), the flux distributions with no extinction will be
applicable for much of the time.  Nonetheless, we want to consider how
much extinction can change the radiation fields.  In order to consider
the effects of extinction on the expected radiation fields, we need to
specify the distribution of gas (and dust), as well as the opacity at
UV wavelengths. For the sake of definiteness, we take the gas
distribution to follow a Hernquist profile (Hernquist 1990) so that
the density is given by
\be
\rho = {\rho_0 \over \xi (1 + \xi)^3} \, , 
\label{density} 
\ee
where $\xi = r/r_s$ and $r_s$ is the scale radius of the 
profile. Here we take $r_s$ = $\rclust$ so that the density 
distribution has the approximate form $\rho \propto 1/r$ 
for radii within the cluster itself. The form of equation 
(\ref{density}) allows the density and its corresponding 
gravitational potential to match smoothly onto the background 
of the molecular cloud, but otherwise plays no role. The 
density scale $\rho_0$ is determined by the specification 
of the gas content of the cluster. For example, if the 
star formation efficiency $\epsilon$ = 1/3, so that two 
thirds of the mass within $\rclust = r_s$ is made of gas, 
then $\rho_0$ = $4 N \langle M_\ast \rangle / \pi r_s^3$, 
where $\langle M_\ast \rangle$ is the mean stellar mass of 
the population. 

With the density specified by equation (\ref{density}), 
the column density is given by the integral 
\be
\ncolumn (r) = {1 \over \mubar} \int_{r_{1}}^{r} \rho(r) dr = 
{\rho_0 r_s \over \mubar} 
\int_{\xi_1}^{\xi} {d\xi \over \xi (1 + \xi)^3 } \, , 
\label{colintegral} 
\ee
where $\mubar$ is the mean mass of the particles and where $r_{1}$ is
an inner cutoff radius. In most clusters, as assumed herein, the most
massive star lies at the cluster center and will evacuate its
immediate vicinity and produce an inner cutoff radius.  Here we set
the inner cutoff radius by assuming that the evacuated cavity
originally contained a mass $QM_{*max}$, leading to an inner radius
\be
r_1 = \left({Q M_{*max}\over 2 \pi\rho r_s^3}\right)^{1/2}\,,
\ee
where we use $Q$ = 3 and 10 to define two choices of $r_1$. 

The integral from equation (\ref{colintegral}) can be evaluated 
to obtain the result 
\be
\ncolumn = { \rho_0 r_s \over \mubar } \left\{ 
\ln \left[ {\xi (1 + \xi_{1} \over \xi_{1} (1 + \xi) } \right] 
+ {1 \over 2 (1 + \xi)^2} - {1 \over 2 (1 + \xi_{1})^2} 
+ {1 \over 1 + \xi} - {1 \over 1 + \xi_{1} } \right\} \, , 
\label{column} 
\ee 
where $\xi = r/r_s$ and $\xi_{1} = r_{1} /r_s$. 

Next we want to obtain a general assessment of the effects of
extinction on the distribution of the UV radiation fields. We assume
that all of the UV-generating stars lie within the spherical cavity
defined above.  The target systems are distributed according to a
$\rho_\ast \sim 1/r$ distribution within the radial range $0 \le r \le
R_c (N)$, i.e., the target stars can orbit through the evacuated
central cavity and experience no extinction (even though they do not
form there).  Each radial position has both an associated flux and an
associated column density $N(r)$ as defined by equation (\ref{column}).  
The column density can be converted into an optical depth through the 
relation
\be 
\tau_{UV} = \sigma_{UV} \ncolumn \, , 
\ee
where the cross sections are given by $\sigma_{FUV}$ = $10^{-21}$ cm$^2$  
and $\sigma_{EUV}$ = $2 \times 10^{-21}$ cm$^2$. 

Note that actual embedded clusters will not necessarily have smooth
distributions of column density and hence extinction.  The dynamic
nature of the star formation process (e.g., winds, jets, and
outflows), and the cloud formation process itself, lead to a highly
clumpy geometry with some less obscured lines of sight.  In such a
system, the degree of penetration of UV radiation, and the
corresponding photo-ionization rates, are much larger than in the case
of uniform density distributions (Bethell et al. 2007). More detailed
radiative transfer models of this type should be performed for these
cluster environments.  One should also keep in mind that the gas only
resides in the cluster for a relatively short time (about 3--4 Myr;
Allen et al. 2007), so the full (unattenuated) UV flux distribution
will be applicable for much of the time.  The treatment presented here
thus represents an upper limit to the effects of extinction on the
distribution of UV fluxes. Nonetheless, as shown next, this effect can
be significant.

\begin{figure}
\figurenum{7a}
{\centerline{\epsscale{0.90} \plotone{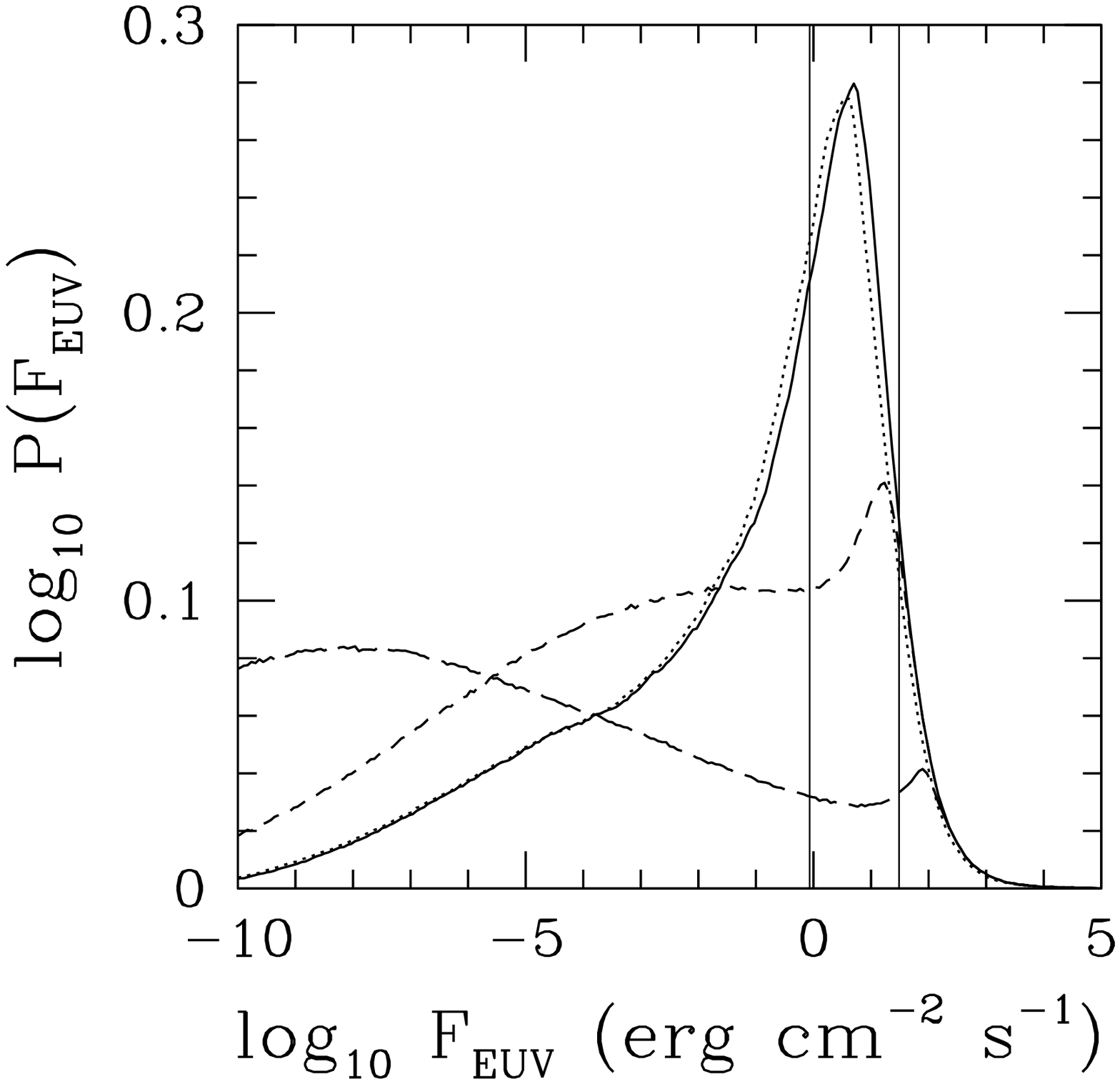} }}
\figcaption{}
\label{fig:flux_euv} 
\end{figure}

\begin{figure}
\figurenum{7a}
{\centerline{\epsscale{0.90} \plotone{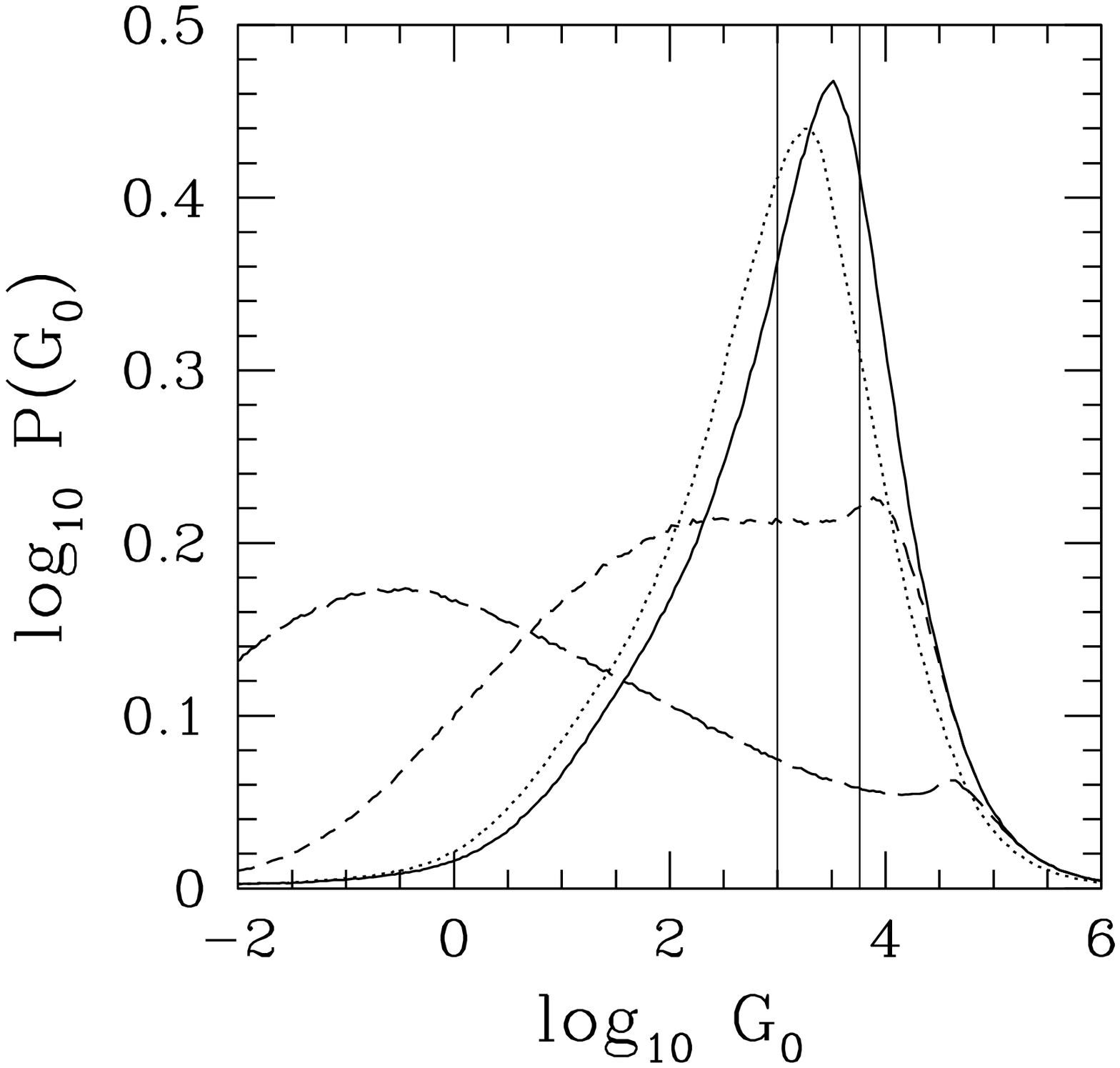} }}
\figcaption{Distribution of EUV flux as a function of EUV flux
$F_{EUV}$ determined by sampling over the Lada \& Lada (2003)
distribution of cluster sizes. This case uses the first stellar IMF
with $(\fone, \Gamma, m_{max})$ = (0.12, 2.35, 100). The solid curve
shows the full distribution; the dotted curve shows the reduced
distribution using only the radiation from the most massive star. 
The dashed curves show the flux distribution with extinction 
included for $Q$ = 3 (long-dashes) and $Q$ = 10 (short-dashes). 
The vertical lines mark the benchmark flux values for which the
background cluster radiation dominates that of the central star (left)
and for which the disk is evaporated in 10 Myr (right). 
b. -- Same as Figure 7a, but for FUV flux.}
\label{fig:flux_fuv} 
\end{figure}

\subsection{Benchmarks}  

One important issue is to determine how often the radiation field
impinging on circumstellar disks is dominated by the background
cluster or by the central star itself.  For any given star and given
background radiation flux, we can determine the radius within the disk
where the UV radiative flux from the central star is equal to that of
the background. For a given central star, let $L_{UV\ast}$ be the
stellar luminosity within a UV band, either FUV or EUV, and let
$F_{UV}$ be the flux in the same UV band from the background
environment of the cluster.  At a given radius $\varpi$ within the
disk, the UV flux contribution from the central star is equal to that
of the background $(F_{UV})$ when
\be
{L_{UV\ast} \over 4 \pi \varpi^2} = F_{UV} \, . 
\label{fluxbound} 
\ee 
The central star dominates at smaller radii and the background cluster
dominates at larger radii.  In order to provide a benchmark for
comparison, we must specify the radius $\varpi$ of interest.  Within
our own solar system, planet formation takes place within 30 AU; in
more general solar systems, the time scale for forming planets
increases with radius and the lifetime of the gas decreases with
radius, so we expect planet formation to become increasingly difficult
for larger values of $\varpi$. We thus adopt $\varpi$ = 30 AU to
evaluate equation (\ref{fluxbound}). The resulting benchmark flux is
thus approximately $F_{EUV}$(bench) $\approx$ 0.86 erg cm$^{-2}$
s$^{-1}$ for the EUV band and $G_0$(bench) $\approx 1000$ for the FUV
band.  Note that throughout this paper, we present FUV fluxes in units
of $G_0$, where $G_0$ = 1 corresponds to the ``standard'' interstellar
value of $1.6 \times 10^{-3}$ erg s$^{-1}$ cm$^{-2}$.

One of the most important effects of the background radiation fields
is to drive photoevaporation from circumstellar disks, thereby leading
to loss of planet-forming potential. We thus want to determine what
part of the expected distribution of UV flux will lead to substantial
mass loss.  For the sake of definiteness, we find the flux required to
evaporate a $M_d$ = 0.05 $M_\odot$ disk over at time scale of 10 Myr.
The expected mass loss rate $\dot M$ from a circumstellar disk exposed
to EUV radiation can be written in the form
\be
{\dot M} \approx 10^{-8} M_\odot {\rm yr}^{-1} \left( 
{F_{EUV} \over 130 \, {\rm erg} \, {\rm cm}^{2} \, {\rm s}^{-1} } 
\right)^{1/2} \left( {r_d \over 30 {\rm AU}} \right)^{3/2} \, , 
\ee 
where $r_d$ is the disk radius (Hollenbach et al. 2000). Note that a
mass loss rate of $10^{-8}$ $M_\odot$ yr$^{-1}$ will evaporate a
typical planet-forming disk with mass $M_d$ = 0.05 $M_\odot$ in only
about 5 Myr.  As a result, EUV fluxes of order $F_{EUV} \sim 30$ erg
cm$^{-2}$ s$^{-1}$ can evaporate this type of disk in 10 Myr and can 
thereby compromise the planet formation process.

For FUV radiation, the detailed models indicate that an external
radiation field of $G_0$ = 3000 will evaporate a disk around a 1.0
$M_\odot$ star down to 36 AU over a time of 10 Myr, where the assumed
disk mass $M_d$ = 0.05 $M_\odot$ (Adams et al. 2004). Interpolating
between the published models for $G_0$ = 3000 and those for $G_0$ =
30,000, we estimate that a radiation flux of $G_0$ = 5800 will
evaporate the disk down to 30 AU in 10 Myr. Note that this benchmark
flux is only about 6 times larger than the flux required to dominate
the stellar FUV flux (see above), i.e., a relatively modest increase
in the FUV radiation environment can lead to a significant effect on 
forming solar systems.  As noted in \S 2, the use of zero-age
main sequence luminosity values is not entirely appropriate in this
context given that the most massive stars (those with $m \sim 100$) 
burn their hydrogen in less than 10 million years.  However, stellar
evolution results in two competing effects for the total cluster
luminosity, since stars get brighter as they age, but the most massive
stars have relatively short lifetimes.  Our results are therefore
expected to be representative of the actual values.

\begin{figure}
\figurenum{8a}
{\centerline{\epsscale{0.90} \plotone{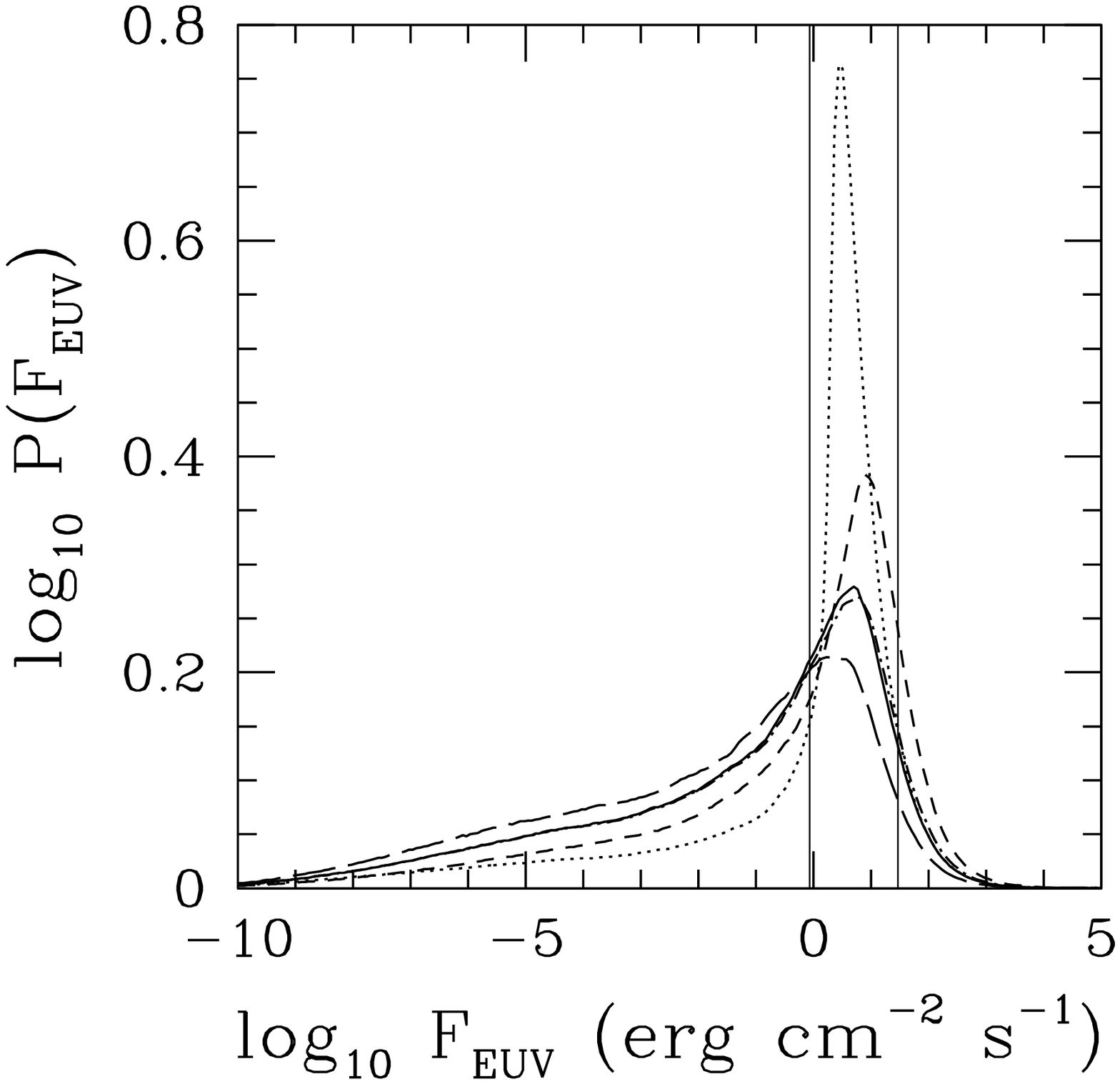} }}
\figcaption{} 
\label{fig:composite_euv} 
\end{figure}

\begin{figure}
\figurenum{8a}
{\centerline{\epsscale{0.90} \plotone{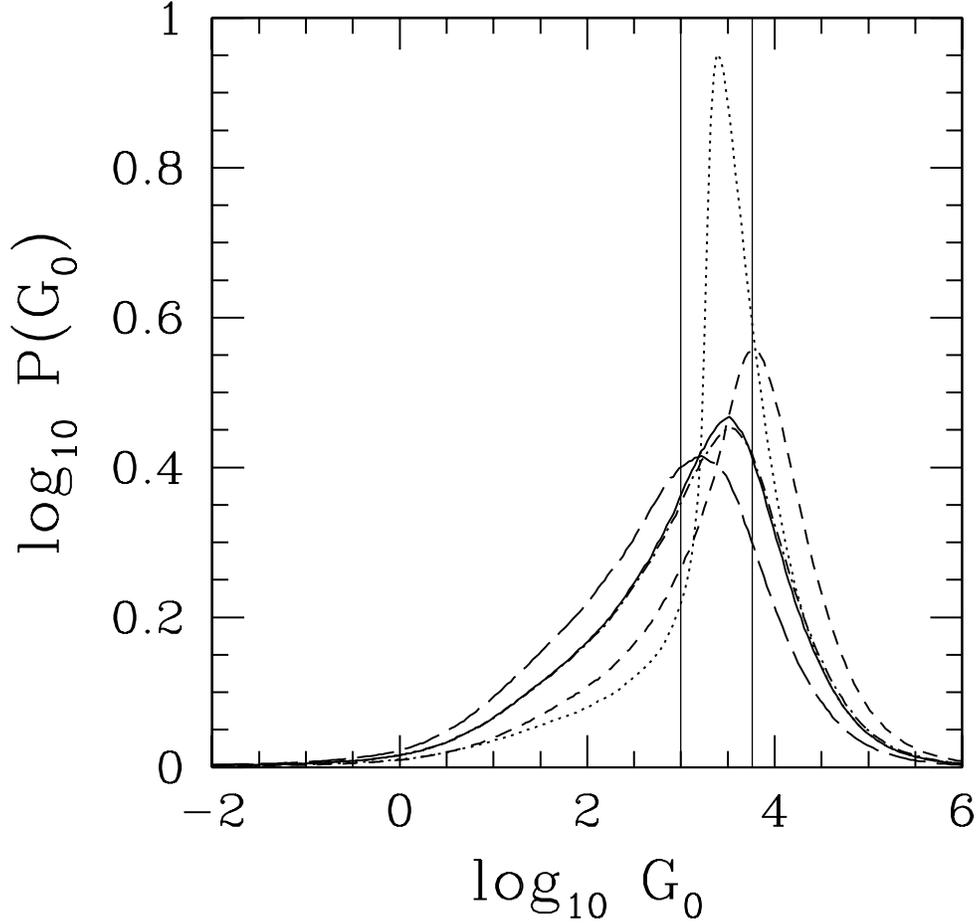} }}
\figcaption{EUV flux distributions for different stellar IMFs and
cluster size distributions.  For the standard cluster size
distribution in the solar neighborhood, the resulting flux
distributions are shown for stellar IMFs with parameters $(\Gamma,
m_{max})$ given by (2.35, 100) (solid curve) (2.1, 100) (short dashed
curve), (2.5, 100) (long dashed curve), and (2.35, 120) (dot-dashed
curve). The dotted curve shows the flux distribution for the standard IMF 
(2.35, 100) and the extended cluster size distribution.  
The vertical lines mark the benchmark flux values for which the
background cluster radiation dominates that of the central star (left)
and for which the disk is evaporated in 10 Myr (right). 
b. -- Same as Figure 8a, but for FUV flux.} 
\label{fig:composite_fuv} 
\end{figure}

\subsection{Composite Flux Distributions} 

Figure 7 shows the flux distributions for the expected cluster
sample. Note that these distributions are determined by first sampling
the cluster size distribution (for the solar neighborhood), then
sampling the standard stellar IMF $N$ times for a cluster of given
size $N$, and finally sampling the radial positions according to a
$\rho \sim 1/r$ distribution. This process is repeated to build up the
distributions shown herein.  The EUV distributions are shown in Figure
7a, and the FUV distributions are shown in Figure 7b. For both of
these cases, the solid curve shows the full distribution for the case
of no extinction; the dashed curve shows the distribution with
extinction included. Also shown in the Figures are the corresponding
distributions for the radiation produced by the most massive star in
the cluster alone (with no extinction).  These latter distributions
are almost identical to the full distributions; this finding indicates
that the radiation fields are dominated by the most massive star in
the cluster for the ensemble of clusters represented in the solar
neighborhood. The dashed curves in Figure 7 show the effects of
including extinction, for the two cases $Q$ = 3 (long dashes) and $Q$
= 10 (short dashes).

The vertical lines in Figure 7 depict the benchmark flux values
defined in the previous section, i.e., the values for which the
background cluster radiation dominates that of the central star (at
disk radius 30 AU) and the values for which the disk is evaporated in
10 Myr (for a disk with mass $M_d$ = 0.05 $M_\odot$ and radius $r_d$ =
30 AU). The former benchmark fluxes are smaller than the latter for
both the EUV and FUV distributions.  For the case of EUV radiation, 42
percent of the distribution has the radiation field dominated by the
background cluster, but only 7 percent of the distribution is exposed
to enough EUV radiation to evaporate the disks.  For the case of FUV
radiation, effect is somewhat larger, with 58 percent of the
distribution being dominated by the background cluster and 25 percent
of the distribution exposed to enough FUV radiation to drive
substantial disk evaporation.  These results thus show that disk
evaporation tends to be dominated by FUV radiation, rather than EUV
radiation, in agreement with previous claims (Hollenbach et al. 2000;
Adams et al. 2004). Furthermore, the dominant source of (EUV)
ionization in circumstellar disks (as required for the
magneto-rotational instability, MRI, for example) is usually the
central star rather than the background (note that cosmic rays and
X-rays also provide ionization and thereby affect MRI).

Figure 8 shows the effects of varying the stellar IMF and the assumed
cluster size distribution on the resulting composite UV flux
distributions. Results are shown for both EUV radiation (Fig. 8a) and
FUV radiation (Fig. 8b). No extinction has been included in the
construction of these distributions. For each case, four of the curves
show the distributions for the four different stellar IMFs used in
this paper.  The upper mass cutoff has relatively little effect on the
composite flux distribution. This finding indicates that the stellar
IMF is sufficiently steep so that stars with the highest masses do not
dominate the UV radiation output. This result is consistent with the
distributions shown in Figure \ref{fig:uvmass}, which indicate that
the UV contribution peaks at masses of 20 -- 40 $M_\odot$.  The slope
of the IMF has a larger effect on the flux distributions, with the
expected result: A shallow slope leads to more high mass stars and
shifts the UV flux distributions to higher values (to the right),
whereas a steeper slope works in the opposite direction.

The fifth (dotted) curves in Figure 8 shows the resulting composite
flux distribution for the extended cluster size distribution (and the
standard stellar IMF). The flux distributions are significantly
narrower for the extended cluster size distribution. Notice that the
peak of the distribution does not change substantially. The reason for
this invariance can be understood as follows: First, note that the
cluster size distribution used here corresponds to equal numbers of
stars being found in each decade of cluster size; as a result, the
extended cluster distribution has only about half of its stars in the
large $N$ clusters, i.e., the clusters that are added to the
distribution for the solar neighborhood.  For these (additional) large
$N$ clusters, the IMF sampling is relatively complete so that the
total luminosity of a cluster obeys the scaling $L_{UV}(tot) \sim N
\langle L_{UV} \rangle_\ast$; however, the cluster radius scales
according to equation (\ref{rofn}), and most stars in the cluster
reside at the larger radii, so that $r^2 \sim N$, and hence the flux
$F_{UV} \sim L_{UV}(tot)/4 \pi r^2$ becomes nearly independent of $N$.

\subsection{Orbits} 

The distributions of flux considered above were constructed by
statistically sampling the distribution of radial positions within a
cluster. However, any given solar system will follow particular orbits
through the cluster. In this subsection, we consider the interplay
between orbital motion and radiation exposure of circumstellar disks.

The first task is to find the orbit-averaged radiative flux.  For the
density profile of equation (\ref{density}), the potential is given by
\be 
\Psi = {\Psi_0 \over 1 + \xi}   \, , 
\ee 
where $\xi = r/r_s$ as above and where $\Psi_0$ = $2 \pi G \rho_0
r_s^2$ determines the total depth of the potential well. If we define
$M_T$ to be the total mass enclosed within the scale radius $r_s$,
which is taken here to be the cluster radius $R_c(N)$, then $\Psi_0$ =
$4 G M_T/r_s$. Following previous treatments (Adams \& Bloch 2005,
APFM), we define dimensionless energy and angular momentum variables
\be
\epsilon \equiv { |E| \over \Psi_0} \qquad {\rm and} \qquad 
q \equiv {J^2 \over 2 \Psi_0 r_s^2} \, , 
\ee 
where $E$ and $J$ are the (physical) specific energy and specific
angular momentum of the orbit. As shown in APFM, the radiation flux 
averaged over an orbit is then given by the expression
\be 
\langle F_{UV} \rangle_{orb} = {L_{UV} \over 8 r_s^2} 
{A(q) \epsilon^{3/2} \over \cos^{-1} \sqrt{\epsilon} + 
\sqrt{\epsilon} \sqrt{1 - \epsilon} } \, , 
\label{orbitave} 
\ee 
where $A(q)$ is a slowly varying function of angular momentum and 
is constrained to lie in the range $1 \le A(q) \le \sqrt{2}$. 

Equation (\ref{orbitave}) gives the orbit-averaged flux for a orbit
with given energy $\epsilon$ and angular momentum $q$. The stellar
dynamics of the cluster determines the distribution of energy and
angular momentum for the cluster members (e.g., Binney \& Tremaine
1987; hereafter BT87). In particular, for given assumptions about the
velocity distribution, one can find the relationship between the
distribution function, the differential energy distribution, and the
density profile of the cluster.  We assume an isotropic velocity 
distribution and a density profile form $\rho \propto 1/r$; note that
this density profile is consistent with our N-body simulations of
young embedded clusters (APFM). For this case, the differential energy
distribution --- the probability distribution for orbital energies ---  
takes the form 
\be 
h(\epsilon) = {d P_m \over d \epsilon} = 
{2 \over (1 - \epsilon_0)^2 } (1 - \epsilon) \, , 
\label{enerdist} 
\ee
which is normalized for the range of dimensionless energies
$\epsilon_0 \le \epsilon \le 1$. In these systems, stellar orbits are
not highly populated for low energies, those well beyond the starting
cluster radius, so we truncate the distribution at some energy scale
$\epsilon_0$. For example, the energy corresponding to a radial orbit
with its outer turning point at twice the nominal cluster radius has
$\epsilon$ = 1/3, which thus defines a representative value. Here we
use $\epsilon_0$ = 1/2, 1/3, and 1/4 to sample the possible values.

\begin{figure}
\figurenum{9a}
{\centerline{\epsscale{0.90} \plotone{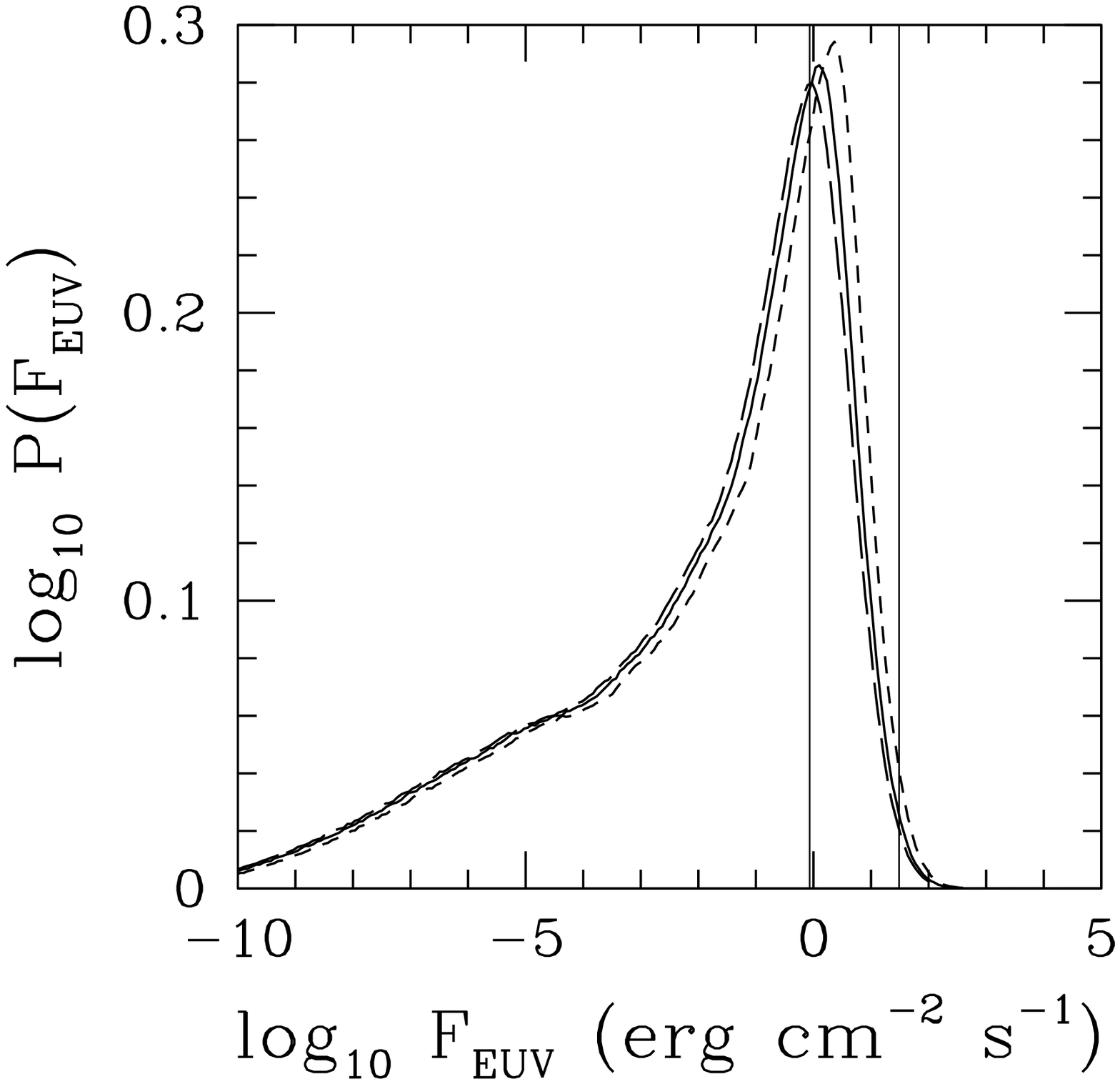} }}
\figcaption{}
\label{fig:orbitflux_euv}
\end{figure}

\begin{figure}
\figurenum{9a}
{\centerline{\epsscale{0.90} \plotone{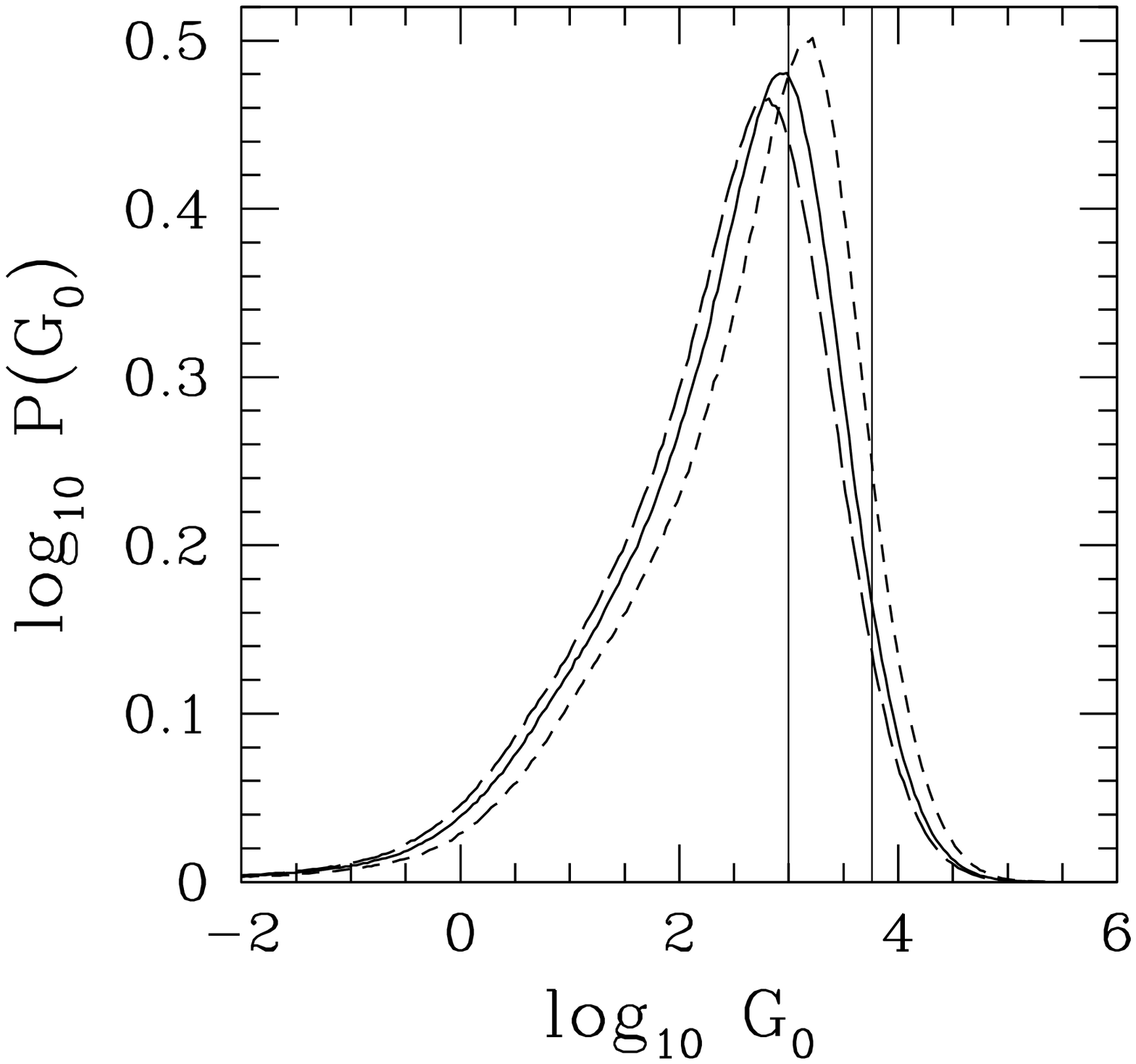} }}
\figcaption{Distribution of orbit-averaged EUV flux. (Compare with
Fig. 7a).  This case uses the standard stellar IMF with $(\fone,
\Gamma, m_{max})$ = (0.12, 2.35, 100) and the Lada \& Lada (2003)
distribution of cluster sizes.  The solid curve shows the full
distribution with $\epsilon_0$ = 1/3; the other two distributions use
$\epsilon_0$ = 1/2 (short dashes) and $\epsilon_0$ = 1/4 (long
dashes).  The vertical lines mark the benchmark flux values for which
the background cluster radiation dominates that of the central star
(left) and for which the disk is evaporated in 10 Myr (right). 
b. -- Same as Figure 9a, but for FUV flux. }
\label{fig:orbitflux_fuv}
\end{figure}

The distribution of energies within the cluster (eq.
[\ref{enerdist}]), in conjunction with the orbit-averaged flux for a
given energy (eq. [\ref{orbitave}]), define a distribution of fluxes
within a given cluster (with given UV luminosity $L_{UV}$, which can
be taken to be either EUV or FUV). Note that the dependence of the
orbit-averaged flux on angular momentum ($q$) is weak, and that the
variations tend to average out over the ensemble of possible values.
We thus need to adopt an intermediate value for $A(q)$ in equation
(\ref{orbitave}). The convolution of equations (\ref{orbitave}) and
(\ref{enerdist}) shows that the resulting distribution of
orbit-averaged flux is weighted toward the lowest flux values, those
corresponding to the lowest energies $\epsilon$, or, equivalently, the
outer parts of the cluster. This result makes sense because the mass
profile $M (r) \sim r^2$, so most of the stars must spend most of
their time in the outer realm of the cluster system.

Figure 9 shows the distributions of orbit-averaged fluxes for both the
EUV (Fig. 9a) and FUV (Fig. 9b) radiation bands. These distributions
are thus the analogs of those shown in Figure 7 (where Fig. 7 was
constructed by sampling in radial position, and Fig. 9 was constructed
by sampling in energy).  In both panels of Figure 9, the solid curve
shows the flux distribution with no extinction, the standard stellar
IMF, the cluster distribution observed in the solar neighborhood, and
the limiting energy value $\epsilon_0$ = 1/3. The other two curves
show the distributions for other choices of the outer boundary
condition: $\epsilon_0$ = 1/2 (short-dashes) and $\epsilon_0$ = 1/4
(long-dashes).  Finally, the two vertical lines are the benchmark
values defined in \S 4.2. The line on the left corresponds to the flux
for which the radiation from the central star is equal to that of the
cluster background at a disk radius of 30 AU; the line on the right
marks the flux for which disks surrounding 1.0 $M_\odot$ stars are
expected to be evaporated over 10 Myr (for assumed disk radius $r_d$ =
30 AU and mass $M_d$ = 0.05 $M_\odot$). Comparison of the
orbit-averaged flux distribution (Fig. 9) with the flux distribution
obtained using the distribution of radial positions (Fig. 7) shows
that net effect of orbit-averaging makes the flux distribution
somewhat narrower and shifts the peak of the distribution to smaller
values; the size of the shift is about a factor of $\sim$3.
Note that the difference between the flux distributions shown in
Figures 7 and 9 represents an upper limit to the effects of orbit
averaging. In real clusters, star-star scattering events and orbital
instabilities will act to move stars from one orbit to another, so
that they will explore many orbits (many values of $\epsilon$). Over
sufficiently long times, solar systems will thus experience the full
distribution of fluxes depicted in Figure 7.

\section{CONCLUSION}

\subsection{Distributions of UV Radiation Fields} 

In this paper we have constructed the distributions of UV radiation
fields expected from the observed collection of young embedded groups
and clusters. This compilation includes both the EUV and FUV bands.
The specification of these radiation fields requires three different
types of input distributions: [1] Clusters with a given stellar
membership $N$ display a UV luminosity distribution due to sampling of
the stellar IMF (note that for $N \simless 10^3$, the sampling is
incomplete and the distributions are wider than their expectation
values, as discussed below). [2] The clusters are found over a
distribution $f(N)$ of sizes $N$. We consider here both the observed
distribution of cluster sizes for the solar neighborhood, where $N$
lies in the range $30 \le N \le 2000$, and an extended cluster size
distribution where we extrapolate the local distribution up to
membership sizes $N$ = $10^5$ (Fig. \ref{fig:clustnum}). [3] The stars
reside at a range of locations within a given cluster, so we must
specify the distribution of radial positions $P(r)$ in the cluster in
order to determine the distribution of radiation flux; alternatively,
we can sample the orbital energy distribution function $h(\epsilon)$
and thereby obtain the orbit-averaged radiation fluxes. These
probability distributions [$f(N)$, $P(L_{UV})$; $P(r)$ or
$h(\epsilon)$] jointly determine the composite distribution of fluxes
that affect the ensemble of forming solar systems. The first two
distributions [$f(N)$, $P(L_{UV}$] are sufficient to determine the
composite distributions of UV luminosities.

Our results show that the distributions of UV radiation are
qualitatively different for small clusters and large clusters, where
the boundary between the two regimes lies in the range $N$ = 1000 --
2000 (note that the boundary is not perfectly sharp).  The
determination of the radiation fields in the smaller clusters is
dominated by incomplete sampling of the stellar IMF, so the resulting
distributions are wider than their expectation values and show
substantial departures from a gaussian form.  In contrast, larger
clusters contain enough stars to sample the IMF, so that their UV
distributions are close to gaussian, with widths less than their 
expectation values. According to this criterion, the entire ensemble
of clusters in the solar neighborhood falls in the range of ``small''
clusters, where UV radiation fields are subject to incomplete sampling
effects.

The effects of the stellar IMF on the radiation distributions are
modest. The upper mass cutoff has little effect, essentially because
the IMF is a steeply declining function of stellar mass, so that stars
in the mass range 20 -- 40 $M_\odot$ provide the most UV radiation
(Fig. \ref{fig:uvmass}). In other words, although more massive stars
produce more UV radiation per object, their total contribution is
diminished due to their rarity.  Shallower slopes for the IMF allow
for greater numbers of massive stars and result in a shift in the
radiation field distribution to higher values (Fig. 8), whereas
steeper IMF slopes act in the opposite direction. The extended cluster
distribution produces a composite UV flux distribution that is
narrower than that of the solar neighborhood, but retains
approximately the same peak value (because the larger cluster radii
compensate for larger cluster UV luminosity as $N$ increases).

We have compared the composite flux distributions using both
distributions of radial positions (Fig. 7) and distributions of
orbital energies (Fig. 9). The first case presents the entire
distribution of radiative fluxes provided by the cluster environment.
The second case presents the distribution of orbit-averaged fluxes
that would be experienced by solar systems as they move through the
cluster in the absence of any interactions (which change the orbits).
The two distributions are similar, with the orbit-avereged flux
distribution being somewhat narrower and shifted to lower values.
The statistical measures for the flux distributions 
shown in Figures 7 -- 9 are summarized in Table 2 (EUV) and
Table 3 (FUV) below.  The first column identifies the distribution
and gives the corresponding figure and line type in parentheses.  
The remaining columns specify the mean, median, peak, and width
of the distributions in terms of the log($F_{UV}$) values
used in the horizontal axis of each corresponding figure.  

\begin{table}
\begin{center}
\caption{\bf Measures for EUV Flux Distributions (log$_{10}$ F$_{EUV}$)} 
\medskip 
\begin{tabular}{lcccc} 
\hline 
\hline 
Distribution (figure and line type)
& Mean & Median &  
Peak & Width \\ 
\hline
Standard (Fig. 7 - solid)& -1.33 & -0.53  & 0.70 & 2.65 \\
Largest Star Only (Fig. 7 - dotted) & -1.43 & -0.60 & 0.57 & 2.64 \\
Extinction with Q = 3 (Fig. 7 - long dash) & -6.95 & -6.60  & -8.0 & 4.10 \\
Extinction with Q = 10 (Fig. 7 - short dash) & -2.89 & -2.53  & 1.23 & 3.36 \\
IMF [2.1, 100] (Fig. 8 - short dash)  & -0.56 & 0.27  & 0.90 & 2.44 \\
IMF [2.5, 100] (Fig. 8 - long dash)  & -1.91 & -1.20  & 0.37 & 2.75 \\
IMF [2.35, 120] (Fig. 8 - dot dash)  & -1.28 & -0.47 &  0.77 & 2.66 \\
extended cluster (Fig. 8 - dotted)  & -0.37 & 0.40  & 0.43 & 2.29 \\
$\epsilon_0 = 1/2$ (Fig. 9 - short dash) & -1.79 & -0.93  & 0.37 & 2.62 \\
$\epsilon_0 = 1/3$  (Fig. 9 - solid) & -1.98 & -1.13  & 0.10  & 2.62 \\
$\epsilon_0 = 1/4$ (Fig. 9 - long dash) & -2.08 & -1.27  & -0.03 & 2.63 \\
\hline 
\hline 
\hline 
\end{tabular}
\end{center} 
\end{table} 

\begin{table}
\begin{center}
\caption{\bf Measures for FUV Flux Distributions (log$_{10} \,G_0$)} 
\medskip 
\begin{tabular}{lcccc} 
\hline 
\hline 
Distribution (figure and line type)
& Mean & Median &  
Peak & Width \\ 
\hline
Standard (Fig. 7 - solid)& 3.06 & 3.20  & 3.52 & 1.13 \\
Largest Star Only (Fig. 7 - dotted)  & 2.84 & 2.97 & 3.28 & 1.14 \\
Extinction with Q = 3 (Fig. 7 - long dash)  & 0.22 & 0.43  & -0.45 & 2.05 \\
Extinction with Q = 10 (Fig. 7 - short dash) & 2.25 & 2.37  & 3.82 & 1.61 \\
IMF [2.1, 100] (Fig. 8 - short dash)  & 3.43  & 3.60  & 3.82 & 1.07 \\
IMF [2.5, 100] (Fig. 8 - long dash)  & 2.77  & 2.90  & 3.22 & 1.15 \\
IMF [2.35, 120] (Fig. 8 - dot dash)   & 3.07  & 3.20  & 3.55 & 1.14 \\
extended cluster (Fig. 8 - dotted)  & 3.43 & 3.47  & 3.38 & 0.97  \\
$\epsilon_0 = 1/2$ (Fig. 9 - short dash)  & 2.58  & 2.77  & 3.22 & 1.05 \\
$\epsilon_0 = 1/3$  (Fig. 9 - solid)  & 2.38  & 2.57  & 2.98 & 1.06 \\
$\epsilon_0 = 1/4$ (Fig. 9 - long dash)  & 2.27 & 2.47  & 2.82 & 1.07 \\
\hline 
\hline 
\hline 
\end{tabular}
\end{center} 
\end{table}

\subsection{Implications for Star and Planet Formation} 

The composite UV flux distributions can be used to provide estimates
for the percentages of forming solar systems that have their radiation
exposure dominated by the background environment of the cluster
(versus the central star). These percentages depend on the mass of the
central star and the radius of interest within the disk. For the case
of solar type stars and disk radii of 30 AU, we find that 42 percent
of the population will have their EUV exposure dominated by the
background, compared to 58 percent for FUV radiation.  Note that the
vast majority of the stellar population has mass smaller than 1.0
$M_\odot$, so the overall percentage of solar systems with radiation
dominated by the background will be higher.

Another way to gauge the importance of these background radiation
fields is to determine the percentage of disks that will be destroyed
by UV radiation before planet formation can take place. For example,
we can consider planet formation to be compromised when the disk
evaporation time becomes less than 10 Myr at a disk radius of 30
AU. For this case, and for the distribution of cluster sizes in the
solar neighborhood, we find that 25 percent of the disk population
loses some of their planet forming potential due to FUV radiation from
the background cluster, whereas only 7 percent of the disk population
is compromised by EUV radiation. This latter result is consistent with
previous claims (and assumptions) about the relative importance of FUV
radiation over EUV radiation in these systems (e.g., Hollenbach et al.
2000; Adams et al. 2004). Again, most stars have smaller masses and
their accompanying disks will be more easily destroyed. For red dwarfs
with $M_\ast = 0.25 M_\odot$, for example, 25 percent of the disk
population will be evaporated down to a radius of $\sim8$ AU, which is
much smaller than the 30 AU benchmark, and will effectively shut down
giant planet formation (see also Laughlin et al. 2004). We note that
additional photoevaporation models of evaporating disks must be done
to provide further quantification of this issue.

Instead of quantifying the likelihood of disk photoevaporation using
the composite flux distribution for the entire cluster sample, we can
consider the loss of circumstellar disk gas as a function of stellar
membership $N$. This question is vital to future and ongoing searches
for extra-solar planets, where clusters are often used as a convenient
means of obtaining a well-defined sample of target stars at known
distances.  The results of this paper show that the radiation fields
produced by clusters with smaller $N$ have much more variation from
cluster to cluster than their larger $N$ counterparts. However, the
mean flux for a cluster is surprisingly insensitive to cluster size
$N$. As shown in \S 4.3, for sufficiently large $N$ the typical UV
flux becomes nearly independent of $N$: In this regime, the total
luminosity $L_{UV} \propto N$, the cluster radius scales like $R_c^2
\propto N$, and most of the stars reside in the outer parts of the
cluster where $r \sim R_c$. As a result, the typical background flux
$F_{UV} \propto L_{UV}/r^2$ becomes nearly independent of stellar
membership size $N$.

Next we note that evaporation tends to remove gas from the outer parts
of circumstellar disks, whereas disk accretion drains gaseous material
from the inner disk. Taken together, these two processes combine to
set the total disk lifetime.  As shown herein, disk evaporation from
FUV radiations dominates over that of EUV radiation for the expected
distributions. Further, the ``typical'' cluster environment provides
enough radiation to evaporate a disk associated with a solar type star
down to a radius of $\sim 30$ AU in 10 Myr. As shown in previous work
(e.g., Clarke et al. 2001; Adams et al. 2004), viscous disk accretion
with viscosity parameter $\alpha \sim 10^{-3}$ results in a disk
lifetime of about 10 Myr for these disk radii (30 AU). As a result,
disk lifetimes of this order of magnitude are expected in cluster
environments, where, indeed, such disk lifetimes have been observed
(Haisch et al. 2001).  One prediction of this work is that disks will
survive longer in the outer parts of clusters, provided that they do
not primarily live on radial orbits. Some observational work on this
issue has been carried out, and suggests that the spatial positions of
circumstellar disks are anti-correlated with the locations of the
massive stars (Guarcello et al. 2007; Balog et al. 2007), 
but more work along these lines
should be done.

EUV leads to ionization (by definition), which has an impact on the
efficacy of MRI as a source of disk viscosity. In the absence of any
background sources of radiation, beyond the EUV flux from the central
star and the background of cosmic rays, circumstellar disks are
expected to have ``dead zones'', regions where the ionization levels
are so low that MRI cannot operate (e.g., Gammie 1996). As shown in \S
4, when compiled over the observed distribution of clusters in the
solar neighborhood, the composite EUV flux distribution shows that 42
percent of the disk population is exposed to significant ionizing
radiation from the environment of their birth clusters, in addition to
that received from the central stars. This percentage was calculated
for a fiducial radius of 30 AU; at this location, the vertical extent
of the dead zones will be smaller, and the efficacy of MRI and disk
accretion will be enhanced for $\sim40$ percent of the population. At
smaller disk radii, however, the percentage of disks that receive a
substantial enhancement of ionizing flux from the background cluster
is much smaller; specifically, for the background to dominate at 0.1
AU, the inner edge of the dead zones, the EUV background flux must be
$\sim10^5$ larger than the benchmark value, and the percentage of the
distribution with such large flux values is negligible. We thus
conclude that the ionizing (EUV) radiation produced by the cluster
background is insufficient to eliminate dead zones in circumstellar
disks.

The radiation fields produced by young embedded clusters also have
implications for the birth aggregate of our solar system. Since a
large fraction of the stellar population is formed within clusters, it
is likely that our solar system was born within a cluster of some size
$N$.  Furthermore, the meteorites from our solar system show evidence
of short-lived radioactive isotopes during their early history, and
this enrichment is often ascribed to a nearby supernova explosion,
which must take place within the birth cluster. Using these observed
properties, and others, a number of authors have tried to determine
and/or constrain the birth environment of our Sun (e.g., Williams \&
Gaidos 2007, Ouellette et al. 2007, Zahnle et al. 2007, Gounelle \&
Meibom 2007, Adams \& Laughlin 2001). The radiation fields produced by
the cluster environment provide additional constraints: If a
hypothetical birth cluster produced too much UV radiation, then the
early solar nebula would be evaporated before the giant planets could
form. In other words, the existence of our giant planets --- in
conjunction with disk photoevaporation models --- places limits on the
environment of our solar system during its first 10 Myr. This paper
(Figs. 7,8,9) shows that the expected distribution of clusters provide
FUV radiation fields in the range $G_0$ = 300 -- 30,000, with a
typical value of $G_0$ = 3000.  Over a time span of 10 Myr, this
latter value of the FUV flux will evaporate a disk around a solar type
star down to 36 AU (Adams et al. 2004), thereby leaving enough gas in
the solar nebula for giant planets to form.  Even a more extreme flux
of $G_0$ = 30,000 would only evaporate the disk down to $\sim15$ AU.
As a result, the gas reservoirs for Jupiter and Saturn are always
safe, whereas the gas supply for Uranus and Neptune could be
compromised (but note that these ice giants have little gas).  In
addition to photoevaporation, circumstellar disks can be destroyed (or
disrupted) in their birth clusters by scattering encounters and by ram
pressure stripping (Pfalzner et al. 2006, Olczak et al. 2006, Throop
\& Bally 2005, Kobayashi \& Ida 2001). These effects have been
included in studies of the solar birth cluster, but can and should be
considered more globally (e.g., Adams et al. 2006, Malmberg et
al. 2007).

\subsection{Discussion and Future Work} 

In addition to continued applications of the distributions of UV
radiation fields constructed herein, as outlined above, the
distributions themselves can be improved in several ways. The greatest
uncertainty concerns the distributions of cluster properties,
especially membership size $N$ and radius $R_c(N)$.  For the solar
neighborhood, the recent observational compilations (Lada \& Lada
2003, Porras et al. 2003) provide good working estimates for the
distributions of cluster membership size $N$ (see Fig. 1) and cluster
radius $R_c$ (see eq. [\ref{rofn}]). Beyond the solar neighborhood,
however, no complete observational census has been carried out, and
one must rely on some type of extrapolation. Is this work, we have
used an extended distribution of the form $dN_C/dN \propto N^{-2}$,
where $N_C$ is the number of clusters, so that equal numbers of stars
are formed within each decade of cluster size $N$ [i.e., $N (dN_C/dN)
\propto N^{-1}$].  For clusters with large $N$, the sampling of the
stellar IMF is relatively complete, so that the statistics of the
distribution of luminosity are well-behaved. In order to determine the
distribution of UV fluxes, one needs the distribution of radial
positions within the cluster, including the cluster radius
$R_c$. Since most of the cluster members do not live in the cluster
core, but rather in its outer parts, the distribution of radial sizes
is crucial. In the present formulation, the adopted relation $R_c
\propto N^{1/2}$ (observed in the solar neighborhood) implies that
although clusters with larger $N$ have correspondingly larger
luminosities, they produce almost the same distribution of fluxes
because of their larger radii. On the other hand, if clusters with
larger $N$ do not follow this empirical law (e.g., they could be more
compact), then the distributions of UV fluxes would be shifted toward
higher values. Another possibility is that the population of larger
clusters has mean radii given by equation (\ref{rofn}), but the radii
sample a wide distribution about the mean, so that the more compact
clusters would produce environments with large UV fluxes.

Another related issue is that this work samples the stellar IMF
independently for clusters of all sizes $N$. Current observartions are
consistent with the assumption that the IMF is independent of
environment (e.g., Kroupa 2002, Chabrier 2003, and references therein). 
However, correlations of stellar IMF with cluster size $N$ could
affect the distributions of UV radiation fields calculated here.
Additional observation work should be carried out to determine if any
such correlations exist, and how they are quantified.

In addition to EUV and FUV radiation, young stars also produce copious
amounts of X-ray emission. As a result, cluster environments can also
provide a significant X-ray background radiation field, which can also
affect star and planet formation. In this case, the dominant effect is
ionization, both in circumstellar disks and in pre-stellar cores. In
the case of disks, ionization helps MRI produce disk accretion, and
thereby helps the star formation process. In pre-stellar cores,
ionization leads to greater coupling between gas the magnetic fields,
and thereby slows down star formation. In future work, the construction 
of X-ray backgrounds in clusters, and the effects of this radiation on 
star formation, should be considered. 

\vskip 0.35truein 
\centerline{\bf Acknowledgment} 

We thank Lori Allen, Tom Megeath, Phil Myers, and Eva Proszkow for
useful discussions an an anonymous referee for useful comments that
improved the paper.  This work was supported at the University of
Michigan by the Michigan Center for Theoretical Physics, by the
Astrophysics Theory Program (NNG04GK56G0), and by the Spitzer Space
Telescope Theoretical Research Program (1290776). This work was
supported at Xavier University through the Hauck Foundation.

\end{document}